# Landauer-QFLPS model for mixed Schottky-Ohmic contact two-dimensional transistors

Zhao-Yi Yan,[1,2,†] Zhan Hou,[1,2,†] Kan-Hao Xue,[3,4,*] Tian Lu,[1,2] Ruiting Zhao,[1,2] Junying Xue,[5] Fan Wu,[1,2] Minghao Shao,[1,2] Jianlan Yan,[1,2] Anzhi Yan,[1,2] Zhenze Wang,[1,2] Penghui Shen,[1,2] Mingyue Zhao,[1,2] Xiangshui Miao,[3,4] Zhaoyang Lin,[5] Houfang Liu,[1,2,*] He Tian,[1,2,*] Yi Yang,[1,2,*] Tian-Ling Ren[1,2,*]

[1] *School of Integrated Circuits, Tsinghua University, Beijing 100084, China*

[2] *Beijing National Research Center for Information Science and Technology (BNRist), Tsinghua University, Beijing 100084, China*

[3] *School of Integrated Circuits, School of Optical and Electronic Information, Huazhong University of Science and Technology, Wuhan 430074, China*

[4] *Hubei Yangtze Memory Laboratories, Wuhan 430205, China*

[5] *Department of Chemistry, Tsinghua University, Beijing 100084, China*

***Corresponding Author**, E-mail: RenTL@tsinghua.edu.cn (T.-L. Ren), yiyang@tsinghua.edu.cn (Y. Yang), tianhe88@tsinghua.edu.cn (H. Tian), houfangliu@tsinghua.edu.cn (H. Liu), xkh@hust.edu.cn (K.-H. Xue)

[†]These authors contributed equally.

## Abstract

Two-dimensional material-based field effect transistors (2DM-FETs) are playing a revolutionary role in electronic devices. However, after years of development, no device model can match the Pao-Sah model for standard silicon-based transistors in terms of physical accuracy and computational efficiency to support large-scale integrated circuit design. One remaining critical obstacle is the contacts of 2DM-FETs. In order to self-consistently include the contact effect in the current model, it is necessary to perform self-consistent calculations, which is a fatal flaw for applications that prioritize efficiency. Alternatively, to improve efficiency, it is necessary to abandon the self-consistency of the model. Here, we report that the Landauer-QFLPS model effectively overcomes the above contradiction, where QFLPS means quasi-Fermi-level phase space theory. By connecting the physical pictures of the contact and the intrinsic channel part, we have successfully derived a drain-source current formula including the contact effect. To verify the model, we prepared transistors based on two typical 2DMs, black phosphorus (BP) and molybdenum disulfide ($MoS_2$), the former having ambipolar transport and the latter showing electron-dominant unipolar transport. The proposed new formula could describe both 2DM-FETs with Schottky or Ohmic contacts. Moreover, compared with traditional methods, the proposed model has the advantages of accuracy and efficiency, especially in describing non-monotonic drain conductance characteristics, because the contact effect is self-consistently and compactly packaged as an exponential term. More importantly, we also examined the model at the circuit level. Here, we fabricated a three-bit threshold inverter quantizer circuit based on ambipolar-BP process and experimentally demonstrated that the model can accurately predict the circuit performance. This industry-benign 2DM-FET model is supposed to be very useful for the development of 2DM-FET-based integrated circuits.

## Introduction

Two-dimensional material-based field-effect transistors (2DM-FETs) have attracted significant interest for their potential to continue Moore's law [1, 2]. Their atomic thickness and dangling-bond-free interface with gate oxide enable high tunability when applied in FET devices [3]. For example, the ambipolar 2DM-FETs, which can transport the electrons and holes simultaneously, are extensively reported for a variety of channel materials, such as graphene (Gr) [4], black phosphorus (BP) [5], tungsten diselenide ($WSe_2$) [6], and molybdenum ditelluride ($MoTe_2$) [7]. Its conduction modes (hole's mode and electron's mode) have opened a brand new way of fabricating highly efficient computational components, which could bring bonus for broad applications, including signal processing [8-10], logic operation[11, 12], communication [13], hardware-security [14], 2DM-based memory [15] and in-memory computing [16, 17]. To explore the system-level applications mentioned above for 2DM-FETs, an experiment-accessible and circuit-deployable physical model that can link the lab data and industrial applications is desired by the electronic design automation (EDA) software.

To this end, computational efficiency and accuracy are equally important. It has been reported that an abnormal nonlinearity that is distinct from the traditional models' prediction will arise from the output curves if significant Schottky barriers are formed at the source/drain contact [5, 18-20]. Unlike silicon, which can form Ohmic contacts through heavy doping, contacts between 2DMs and metals mainly belong to the Schottky category [1]. Due to the contact issue, the models currently developed generally make a trade-off between physical accuracy and computational efficiency, which has become a key bottleneck in developing 2DM-FETs' models applicable for EDA tools. Since carriers pass through a series-connection of "source contact/channel/drain contact," an appropriate modeling would not describe the contact and the channel flows separately, where the former is essentially quantum and the latter remains in the semiclassical or even classical domain in most cases. Therefore, to self-consistently describe the channel current, including contact effects, it should be accomplished by seamlessly connecting two physical pictures. This is a challenging task, and according to traditional views, it would lead to a very complex model.

The previously reported models, represented by the virtual-source model [21-23] and the Landauer model [24-27], completely ignore the drift-diffusion process in the channel. The virtual-source model is an empirical model that affords little information regarding transport mechanisms. The Landauer model is considered accurate in describing coherent transport at the quantum regime, but it completely ignores the possible carrier scattering inside the channel. Recently reported models that simultaneously handle the contacts and channel are still based on the scheme of equivalent circuits. One is the equivalent parasitic resistance (EPR) method, which regards the contacts as parasitic nonlinear resistors and substitutes them into the circuit to solve the resulting circuit equation [21-23, 28-33]. These equations usually have to be iteratively solved, sacrificing the efficiency. The other is the non-self-consistent model, which regards the contact as a device in parallel with the channel, calculates their currents separately, and then takes a weighted average as the equivalent current [34-36]. Despite its remarkable efficiency, this method is an *ad hoc* approach that lacks a sound physical foundation. Among them, although Ref. [34] attempted to introduce the electrostatic relation of the channel when calculating the junction energy bands of the source and drain contacts, a theoretical problem persists that the impact of the contacts on the boundary values of the quasi-Fermi levels (QFLs, or "electrochemical potential") was ignored. In other words, the contact and the channel are still treated independently in that model, and a "post-processing" had to be performed to splice the results, failing to consider the coupling between the contacts and the channel. Other works [37, 38] did incorporate the contact current effect into the FET models, but the time consumption during simulation is a hindrance to their application in practical circuit design.

It turns out that an ideal model solution should comprehensively reflect the transport characteristics of the contacts and channel, while maintaining the computational efficiency that is feasible for circuit level simulation. This is an extremely difficult task and fundamentally requires innovative bottom-up theoretical methods. A breakthrough may be expected through our recently developed quasi-Fermi level phase space (QFLPS) theory [39], that can be used to describe the intrinsic channel very efficiently. This theory rigorously considers the influence of QFL splitting on transport characteristics and self-consistently includes channel electrostatics. The present work, signified by "*Landauer-QFLPS modeling*", further proposes that the Landauer formula describing contact transport can be cogently connected to the QFLPS model describing channel transport. A rather surprising conclusion of the work is that the Landauer formula can be written in QFLPS-model form augmented by a barrier attenuation factor based on 2D carrier state density, intimately connecting the physics of the two regions. With this theoretical result, an efficient expression of the current that does not depend on intermediate variables can be derived.

Based on the established model, three verification sections follow thereby. In "*Model verification: BP-FET*", we use BP, a typical ambipolar 2DM, as an example, to demonstrate the model's capability of mastering experimental data. Based on standard parameter extraction methods, complete I-V data collected from a dozen of devices have been tested. In "*Tape-out verification: BP-ATIQ circuit*", we tape out a three-bit ambipolar threshold inverter quantizer (ATIQ) circuit with the ambipolar-BP process to demonstrate the model's circuit simulation capability. ATIQ circuit can be used in flash analog-to-digital converters to reduce chip area and power consumption. Here, the study focuses on predicting and optimizing the circuit's performance to demonstrate its value for EDA software. In "*Unipolar version: MoS2-FET verification*", we also verify the model against an $MoS_2$-based FET, which typically exhibits n-type transport, to show that the coverage of the model in unipolar transport as well.

Recent studies have proposed novel contact technologies that can achieve Ohmic contacts [40-42], but their feasibility for industrial production is still an open question due to process-compatibility issues. For example, Bi can suppress the generation of metal-induced gap states (MIGS) due to its semi-metallic band dispersion characteristics, resulting in Ohmic contacts [42]. However, both Bi and Sn are low-melting-point metals, which are incompatible with back-end-of-line (BEOL) processing, hindering the practicality of this technology. 2DM transistors prepared by conventional contact processes with large-scale production potential generally exhibit Schottky characteristics. Hence, it is generally eager to model devices by combining carrier injection and channel transport [2], and this work would like to fill this gap.

## Landauer-QFLPS modeling

The Landauer-QFLPS model is constructed as follows. The channel region is divided into three parts: (i) the source-contact region $[x_s, x_{si}]$, (ii) the intrinsic channel $[x_{si}, x_{di}]$, and (iii) the drain-contact region $[x_{di}, x_d]$. With the gate-source voltage $V_{GS}$ and drain-source voltage $V_{DS}$ (assuming $V_{DS}$> 0 for clarity) applied, the current $I_{DS}$ flows from the drain to the source (as shown in **Fig. 1**a) and includes both electron and hole components, i.e., $I_{DS} = I_e + I_h$. It is assumed that the Schottky barrier at the source blocks the injection of electrons into the channel, but a finite electron flow is present due to the thermal effect or tunneling effect, leading to the electron-injection current $I_{es}$ on the $[x_s, x_{si}]$ interval. At the same time, electron-QFL $\varepsilon_{Fn}$ drops from $\varepsilon_{Fs}$ at $x_s$ to $\varepsilon_{Fni}$ at $x_{si}$ (as shown by the blue dotted line in **Fig. 1**b). Similarly, holes overcoming the Schottky barrier at the drain generates a hole-injection current $I_{hd}$ on the $[x_{di}, x_d]$, interval and the hole-QFL $\varepsilon_{Fp}$'s rising from $\varepsilon_{Fd}$ at $x_d$ to $\varepsilon_{Fpi}$ at $x_{di}$ (as shown by the red dotted line in **Fig. 1**b). The energy band profile assumed here can be generalized when necessary. The internal QFLs, $\varepsilon_{Fni}$ and $\varepsilon_{Fpi}$, are implicitly determined through the conservation laws of electron and hole currents, respectively, i.e., $I_{es} = I_e$ and $I_{hd} = I_h$. Once $\varepsilon_{Fni}$ and $\varepsilon_{Fpi}$ are determined, the conservation quantities $I_e$ and $I_h$ (or $I_{es}$ and $I_{hd}$) can be calculated, and thus $I_{DS}$ can be obtained. However, solving the internal QFLs is cumbersome. A central message of this work is that $I_{DS}$ can be determined without explicitly finding $\varepsilon_{Fni}$ and $\varepsilon_{Fpi}$. Here, the intrinsic channel currents $I_e$ and $I_h$ are already described by the QFLPS model, i.e., the integrals of carrier densities with their QFLs' paths. Therefore, the key proposal is developing QFLPS-model-like forms for the contact currents $I_{es}$ and $I_{hd}$, which is achieved through some intriguing derivation below.

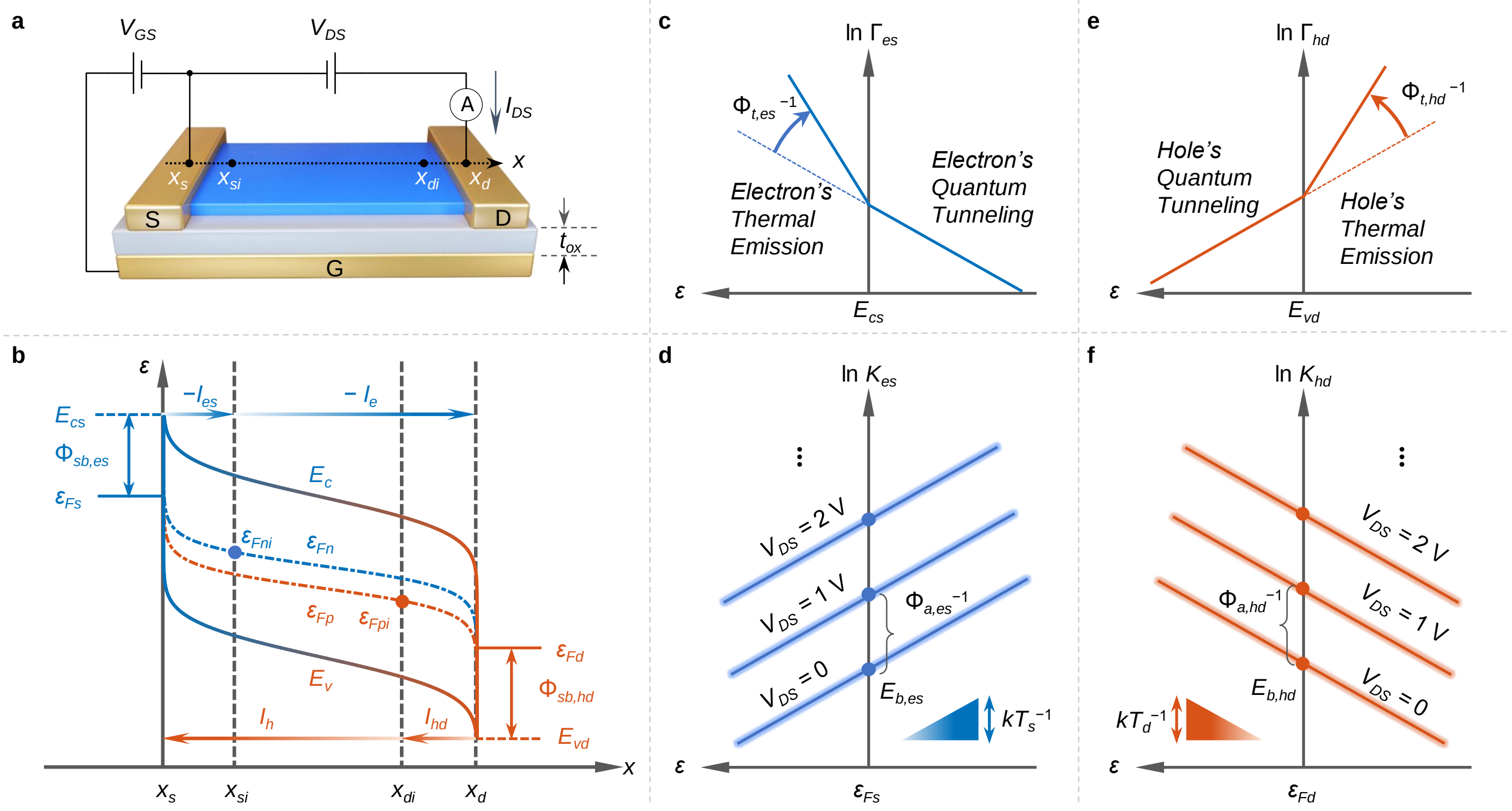

**Fig. 1 | The schematic diagram of the Landauer-QFLPS model principle**. **a**. 2DM-FET device structure and electrical testing schematic; **b**. Energy band diagram of current transport mechanism; **c**. Source electron transmission energy spectrum model; **d**. Source electron collective kinetic energy spectrum model; **e**. Drain conductance hole transmission energy spectrum model; **f**. Drain conductance hole collective kinetic energy spectrum model.

For simplicity, we here focus on the case for electrons, and the derivation for the holes is similar. As the injection of electrons from the source to the channel occurs in an atomically thin space, the relation between the amount of QFL lowering and the resulting current $I_{es}$ should be described by the Landauer formula [24]

$$I_{es} = W\frac{q}{\pi\hbar}\int_{-\infty}^{+\infty}\Gamma_{es}(\varepsilon)M_{es}(\varepsilon)[f(\varepsilon,\varepsilon_{Fs}) - f(\varepsilon,\varepsilon_{Fni})]d\varepsilon \tag{1}$$

where $W$ is the channel width, $q$ is the elementary charge, and $\hbar$ is reduced Planck constant. $\Gamma_{es}(\varepsilon)$ is the transmission energy spectrum of injected electrons at the source, while $M_{es}(\varepsilon)$ is the corresponding density-of-mode (DOM) energy spectrum. Function $f(\varepsilon, \varepsilon_F) = (1 + \exp{(\varepsilon - \varepsilon_F)/kT})^{-1}$ is the Fermi-Dirac distribution function, where $k$ represents the Boltzmann constant, and $T$ represents temperature.

The transmission energy spectrum $\Gamma_{es}(\varepsilon)$ is modeled considering the following fact: for electron energy higher than the source-contact barrier ($\varepsilon > E_{cs}$, where $E_{cs}$ represents the conduction band edge $E_c$ at $x_s$ and the Schottky barrier for source electrons $\Phi_{sb,es} \coloneqq E_{cs} - \varepsilon_{Fs}$, as shown in **Fig. 1**b), the thermal emission mechanism dominates, while for electron energy lower than the barrier ($\varepsilon < E_{cs}$), quantum tunneling dominates. This physical picture (**Fig. 1**c) can be summarized by the following equation:

$$\Gamma_{es}(\varepsilon) = \exp\left(-2\gamma + \Lambda(\varepsilon - E_{cs}\,)/\Phi_{t,es}\right) \tag{2}$$

where $\gamma = (2\hbar^2 q\rho_s/m_e^* m_0 \epsilon_s)^{-1/2}(E_{cs} - \varepsilon)$ is tunneling factor derived from the Wentzel–Kramers–Brillouin (WKB) approximation for the tunneling process (Supplementary Note 1). Here, $m_0$, $m_e^*$, $\rho_s$, and $\epsilon_s$ represent the electron mass, electron's relative effective mass, local effective charge density, and dielectric constant, respectively. $\Lambda(\cdot)$ represents the ramp function, and $\Phi_{t,es}$ represents the thermal-emission energy barrier for the injected electrons from the source.

The DOM function $M_{es}(\varepsilon)$ is given by the formula [25]

$$M_{es}(\varepsilon) = \frac{g_v}{\pi h}\sqrt{2m_{es}^* m_0 K_{es}(\varepsilon)} \tag{3}$$

where $g_v$ is the valley degeneracy, $m_{es}^*$ is the relative effective mass of source electrons, and $K_{es}(\varepsilon)$ is the collective energy spectrum of electrons, taking into account the energy-level occupancy described by the Maxwell-Boltzmann distribution in thermal equilibrium. However, what is of interest is not the thermal equilibrium but the situation where the system is driven away from thermal equilibrium by applied $V_{DS}$. Therefore, the $K_{es}(\varepsilon)$ spectrum should be modified accordingly to lift the collective momentum to a higher level, as shown in **Fig. 1**d. With this acceleration effect, $K_{es}(\varepsilon)$ is written as

$$K_{es}(\varepsilon) = E_{b,es}\exp\left(\frac{qV_{DS}}{\Phi_{a,es}}\right)\exp\left(-\frac{\varepsilon - \varepsilon_{Fs}}{kT_s}\right) \tag{4}$$

where $E_{b,es}$ represents the elemental kinetic energy when $V_{DS} = 0$, and $\Phi_{a,es}$ represents the acceleration barrier for the source electrons, while $T_s$ represents the temperature of the source contact. Based on Eqs. (2)-(4) and 2D density-of-states (DOS) of the channel carriers (introduced in *Method*), the electron-injection current $I_{es}$ at source contact defined by Eq. (1) can be transformed into a QFLPS-like form as (Supplementary Note 1)

$$I_{es} = \exp\left(-\eta_{es} + \frac{qV_{DS}}{\Phi_{a,es}}\right)\frac{W}{L}\int_{\varepsilon_{Fni}}^{\varepsilon_{Fs}} \mu_n n \, d\varepsilon_{Fn} \tag{5}$$

where $\mu_n$ is the electron mobility, $n$ is the 2D-electron density, and $L$ is the channel length. And, the reciprocal of the pre-exponential factor of Eq. (5), i.e., $\exp\left(\eta_{es} - qV_{DS}/\Phi_{a,es}\right)$, is defined as the barrier attenuation factor (BAF) for source electrons with $\eta_{es}$ defined as the contact-current-limiting (CCL) index for source electrons (Supplementary Note 1) and shown to be equal to

$$\eta_{es} = \ln\left[\frac{m_e^* \mu_n/q}{L\frac{\sqrt{2m_{es}^* E_{b,es}}}{8\pi^2\Phi_{t,es}}\exp\left(-\frac{\Phi_{sb,es}}{2kT_s}\right)}\right] \tag{6}$$

The significance of Eq. (5) lies in that we connect the incoherent and coherent transports with the BAF factor. On the one hand, the integral of $n$ in Eq. (5) represents the QFLPS model describing an incoherent transport and arises by integrating the drift-diffusion current along the spatial dimension. On the other hand, $I_{es}$ originates from the coherent transport Eq. (1), where the current is given an integral over energy dimension. Therefore, the united form indicates that although the two current mechanisms are different, they can be unified in the QFL dimension. During this transformation, the unique carrier DOS of 2D materials and the abrupt transmission spectrum near the band edge play a crucial role.

After the electrons are injected from the source into the channel, the source-injection current $I_{es}$ is converted as the intrinsic-channel current $I_e$, which is described by the QFLPS model [39], i.e., the integral of $(W/L)\mu_n n$ over the interval $[\varepsilon_{Fd}, \varepsilon_{Fni}]$. Using the electron current conservation condition $I_e = I_{es}$, the internal QFL $\varepsilon_{Fni}$ can be implicitly eliminated from Eq. (5), which leads to the Landauer-QFLPS formula (Supplementary Note 1)

$$I_e = \frac{1}{1+\exp\left(\eta_{es} - qV_{DS}/\Phi_{a,es}\right)}\frac{W}{L}\int_{\varepsilon_{Fd}}^{\varepsilon_{Fs}} \mu_n n d\varepsilon_{Fn} \tag{7}$$

The pre-factor in Eq. (7) is always less than 1. Hence, it reflects the contact effect included in this model. Apparently, the effect becomes more pronounced with greater CCL index $\eta_{es}$ or the acceleration barrier $\Phi_{a,es}$. Therefore, the model continuously describes the transition from Schottky to Ohmic contact by varying $\eta_{es}$ or $\Phi_{a,es}$ parameters. Since $\eta_{es}$ comprehensively incorporates nearly all the model parameters, the following analysis is mainly focused on it.

Table I Parameter estimations for $\eta_{es}$

| Quantity | Magnitude order |
|---|---|
| $m_e^*$ | $10^{-30}$ kg |
| $m_{es}^*$ | $10^{-30}$ kg |
| $\mu_n$ | $10^{-2}$ m$^2$ V$^{-1}$ s$^{-1}$ |
| $L$ | $10^{-6}$ m |
| $\Phi_{t,es}$ | 1 eV |
| $E_{b,es}$ | 1 eV |

Basically, $\eta_{es}$ measures the relative transport capabilities between the intrinsic channel and source contact according to its expression Eq. (6), where the numerator actually gives the carrier lifetime $\tau_n \coloneqq m_e^* \mu_n/q$ due to the intra-band relaxation in the channel, while the denominator defines an effective source-contact-electron lifetime $\tau_{n,es}$ for the source contact. Significant contact effect occurs if $\tau_{n,es}$ is remarkably shorter than $\tau_n$. With channel materials fixed, properly choosing the contact metal to reduce the barriers $\Phi_{t,es}$ and $\Phi_{sb,es}$ or increase $E_{b,es}$ can extend $\tau_{n,es}$. The channel length $L$ in the denominator actually reminds that an appropriate contact is much important in the short-channel device. We can have a rough idea for $\eta_{es}$'s orders of magnitude by estimating the parameters given in Table I.

With parameters in Table I, we can estimate that $\eta_{es} \sim \Phi_{sb,n}/2kT_s$, which means that $\eta_{es}$ will approach 0 if the Schottky barrier is lowered to a negligible level. In this case, the pre-factor in Eq. (7) will tend to 1, making the contact effect negligible and consistent with our physical intuition of device behavior. Thus, the above analysis confirms the model's validity to a certain degree.

For holes, the physical pictures drawn above need inverting on the energy dimension. For example, for drain-injection holes' the transmission spectrum $\Gamma_{hd}(\varepsilon)$ (**Fig. 1**e), hole's thermal emission mechanism is activated when the energy is lower than the valence band maximum, while its quantum tunneling dominates when the energy is higher than it. For $K_{hd}(\varepsilon)$, the adaptions are similar (**Fig. 1**f). Hence, the Landauer-QFLPS model formula for holes can be derived as (Supplementary Note 2)

$$I_h = \frac{1}{1 + \exp\left(\eta_{hd} - qV_{DS}/\Phi_{a,hd}\right)} \frac{W}{L} \int_{\varepsilon_{Fd}}^{\varepsilon_{Fs}} \mu_p p d\varepsilon_{Fp} \tag{8}$$

where $\mu_p$ is the hole mobility, $p$ is the 2D-hole density, $\Phi_{a,hd}$ is the acceleration barrier for the drain holes, and $\eta_{hd}$ is the hole's CCL index at the drain contact defined as

$$\eta_{hd} = \ln\left[\frac{m_h^* m_0 \, \mu_p/q}{L \frac{\sqrt{2m_{hd}^* m_0 E_{b,hd}}}{8\pi^2 \Phi_{t,hd}} \exp\left(-\frac{\Phi_{sb,hd}}{2kT_d}\right)}\right] \tag{9}$$

where $m_h^*$ is the relative effective mass of holes, $\Phi_{t,hd}$ is the thermal emission barrier for drain holes, $m_{hd}^*$ is the drain holes' relative effective mass, $E_{b,hd}$ is the elemental thermal equilibrium kinetic energy of holes in the drain contact, $T_d$ is the equivalent temperature of the drain contact, and $\Phi_{sb,hd}$ is the corresponding Schottky barrier.

In a word, the Landauer-QFLPS model gives the total current as $I_{DS} = I_e + I_h$, where $I_e$ and $I_h$ are described by Eq. (7) and (8), respectively. They distinguish with the intrinsic QFLPS model by the respective pre-factors to describe the contact effects at the source and drain. It is worth mentioning that this unique pre-factor eliminates the conductivity divergence that occurs in the intrinsic QFLPS model for short-channel limits ($L \rightarrow 0$) and provides a finite value.

The modeling equations for 2D-carrier densities $n$ and $p$ have been described in detail in the previous work [39, 43]. For the self-consistency of this work, these formulae are included in the *Method* for reference. The input parameters required there include the interface trap densities of electrons and holes, $N_{it,e}$ and $N_{it,h}$, and the thermal activation barrier of the channel carrier, $\Phi_t$. Other physical parameters, such as the relative effective masses of carriers, $m_e^*$ and $m_h^*$, can be found in published literature. Therefore, a nine-parameters list $\{\mu_n, N_{it,e}, \mu_p, N_{it,h}, \Phi_t, \Phi_{a,es}, \eta_{es}, \Phi_{a,hd}, \eta_{hd}\}$ needs specified. For practical ambipolar devices, electrons and holes can share a set of model parameters that characterize the contact effect (because ambipolar transport phenomena can hardly be observed if the CCL indices of electrons and holes differ too much), and $\{\Phi_{a,es}, \eta_{es}, \Phi_{a,hd}, \eta_{hd}\}$ can be simplified as $\{\Phi_a, \eta\}$, so a seven-parameters list $\{\mu_n, N_{it,e}, \mu_p, N_{it,h}, \Phi_t, \Phi_a, \eta\}$ can be used instead. Furthermore, for unipolar devices, such as n-type (p-type) devices, $\mu_p$ ($\mu_n$) can be set to 0, and the remaining model variables can be used to describe the electrical characteristics.

In the ideal case of the model assumptions, all model parameters should be constants. However, physical processes in practical devices, such as channel carrier scattering with lattice phonons, carrier-carrier scattering, and Coulomb scattering induced by ionized impurities and defects, result that the mobility is not constant. The local electrical environment should influence the ionization degree of interface traps for electrons and holes. As for the conductivity of the contact, it should also be regulated by the gate-source voltage, as the gate electric field can affect the barrier height. Based on the above analysis, the model parameters should vary with the external bias. Due to the physical processes' complexity, which is constrained by experimental conditions and fluctuations in environmental and material properties, it is not easy to systematically model them. However, thanks to careful consideration along the $V_{DS}$ variation dimension, a simple but efficient model parameterization strategy is proposed here, that is, the model parameters are assumed to only depend on $V_{GS}$ and can be approximated by a universal low-rank function. The coefficients of the low-rank function serve as parameters to be calibrated by experimental optimization. It has been found that linearized Gaussian wavelet functions [44] are well-suited for approximating the low-rank function. Technical details about this part are described in the *Method* section.

## Model verification: BP-FET

BP is a classic 2DM platform demonstrating ambipolar current transport [45]. Unlike the transition-metal dichalcogenides (TMDs) family, BP can maintain a direct bandgap from the single-layer to the multi-layer range. The bandgap can be tuned from 2 eV in the single layer to 0.3 eV in the bulk material, which perfectly bridges the bandgap gap between Gr and TMDs. BP with a few-nanometer thickness obtained by mechanical exfoliation typically has a few hundred milli-electron volts bandgap and a switch ratio of about $10^3$ at a drain-source bias of 0.1 V [45]. Taking advantage of the ambipolar transport characteristics of BP, numerous promising applications have been reported in the literature.

Here, a batch of BP transistors was experimentally prepared by mechanical exfoliation. The device's optical microscope image is shown in **Fig. 2**a. The device adopts a back-gate structure (BG). The channel part is thermally evaporated with electrodes, which can be used as the gate, source, and drain of the transistor, respectively, to apply gate-source voltage $V_{GS}$ and drain-source voltage $V_{DS}$. The fabrication process is described in the *Method* section. Mechanical exfoliation was chosen as the film preparation method mainly for two reasons: (i) the preparation process of mechanical exfoliation is simpler and faster, which is suitable for BP, a material that is readily oxidized by water vapor and facilitates the preservation of its intrinsic properties; (ii) the material obtained by mechanical exfoliation has inherent feature of random dispersion due to the uneven external mechanical stress introduced during the tearing and transfer process. Thus, it can provide device data with rich performance, which can be used for more harsh model testing. As shown in **Fig. 2**b, the BP film in the channel region was characterized by atomic force microscopy (AFM), and the thickness was determined to be approximately 9 nm (±0.7 nm). Transmission electron microscopy (TEM) combined with energy dispersive spectroscopy (EDS) characterizes ~ 2 nm natural phosphorus oxide (POx) layers existing at the interfaces of BP with Ti- and $Al_2O_3$- layers (Supplementary Note 4). Hence, the intrinsic BP thickness is around 5-7 nm and is less than the thickness characterized by AFM, which is normal for BP [46]. After the device fabrication, Raman spectrum of the channel region was measured (**Fig. 2**c), which displayed three distinct and sharp characteristic peaks, indicating that the BP sample did not experience significant degradation during the preparation process.

We sweep the drain-source voltage $V_{DS}$ from 0 to 3 V and the gate-source voltage $V_{GS}$ from −3 V to 3 V to test the BP-FETs. As the result, 15-sets BP-FETs' I-V data are collected. The selected voltage's range covers all possible voltage inputs within the power supply voltage $V_{DD}$ = 3 V to make a full comparison, i.e., $-V_{DD} < V_{GS} < V_{DD}$ and $0 < V_{DS} < V_{DD}$. This test setting should be standard for ambipolar-device's experimental calibrations. However, it was rarely obeyed in practice [47], which limits the persuasiveness of the model's accuracy. A relevant cause is that the threshold voltage of the prepared device usually falls in a non-standard range, so the test voltage has to be shifted. In contrast, the BP-FETs in this article allows performance evaluation in the standard voltage range, and is advantageous for demonstrating the application of the model in circuit simulation, as shown in the next section.

The measured I-V data, their model simulations, and extracted parameters are given in Supplementary Note 5 due to their large data-volume. We note that the extracted $\eta$ parameter exhibits a bell-shaped distribution with the lowest value about 0 and the highest value approaching 4, while most of devices' $\eta$ concentrate on 1-3 interval (**Fig. 2**d). According to our theory, the $\eta$ factor measures the degree of contact effect. The extracted distribution means that devices with nearly ideal Ohmic contact and devices with significant Schottky contact are rare. Most of devices are in an intermediate stage. Actually, both Ohmic and Schottky contacts for 2DM FETs have been experimentally reported [5, 13, 18-20, 38, 48-53]. Experiments within similar materials systems can even exhibit both types of devices [6, 49, 54-57], too. Hence, it is essential to consider the coexistence of Ohmic and Schottky contacts in modeling practical devices.

Specifically, the experimental data and their model simulations of three devices with representative $\eta$ values, Dev#5-6 (with the lowest $\eta$), Dev#1-2 (with medium $\eta$), and Dev#2-3 (with the highest $\eta$), are compared and presented. As shown in **Fig. 2**e-m, they are the transfer (**Fig. 2**e-g), output (**Fig. 2**h-j), and drain conductance (**Fig. 2**k-m) curves of the three devices, respectively. The simulated transfer curves in **Fig. 2**e-g reflect the matching on the global current range, while the output curves (**Fig. 2**h-j) focus on the matching for the on-state current. Their transfer and output curves show a typical ambipolar feature, i.e., the hole current dominates and decreases with increasing $V_{GS}$ when $V_{GS} < 0$, then, the electron current dominates and increases with increasing $V_{GS}$ when $V_{GS} > 0$. The simulation results show that in the complete voltage range, the model simulation results can highly match the experimental test data.

The drain conductance $G_{DS} \coloneqq \partial I_{DS}/\partial V_{DS}$ shows significantly different patterns among the devices, especially where the electron dominates (roughly when $V_{GS} > 1.2$ V). For example, $G_{DS}$ for the typical case of $V_{GS} = 3$ V is studied (**Fig. 2**k-m). The experimental data for the drain conductance curve is obtained from the finite difference approximation (FDA) of the discrete data. Due to the inevitable noise signals contained in the experimental signals, the results obtained by FDA method will be affected by local fluctuations. Varying FDA's range can, to some extent, offset the noise's influence, which leads to the error bars in **Fig. 2**k-m. The device with a smaller $\eta$, Dev#5-6, exhibits a decreasing $G_{DS}$-trend (**Fig. 2**k), which indicates the intrinsic saturation of the electron-dominated drain current for an Ohmic-contact device (**Fig. 2**k). In contrast, $G_{DS}$ of Dev#1-2 and Dev#2-3, undergoes a non-monotonic change (**Fig. 2**l and **Fig. 2**m), which is an important feature caused by Schottky contact. It can be seen that the model can reproduce all these conductivities well within the error range.

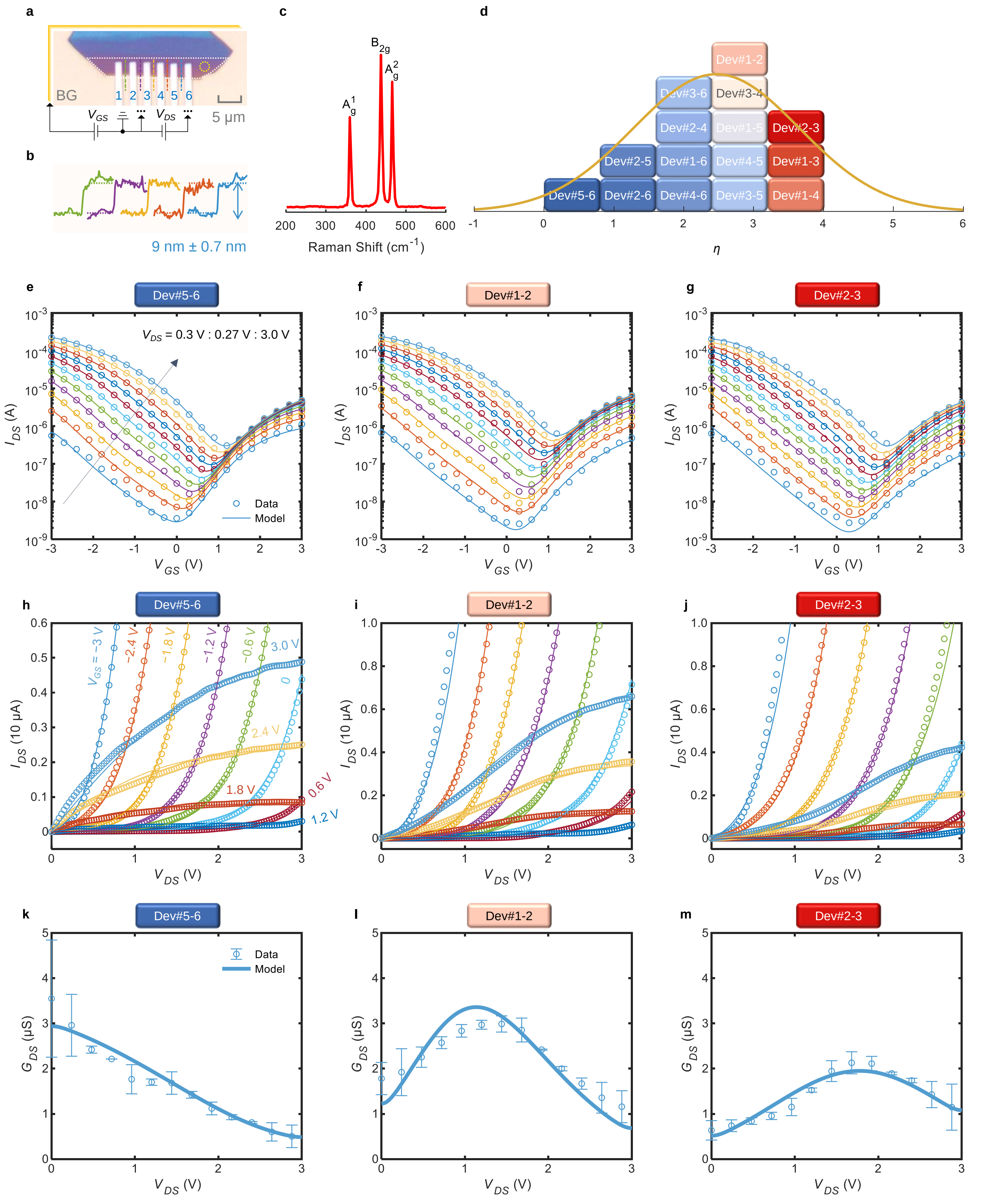

**Fig. 2 | Experimental Verification --- BP-FETs. a.** Optical microscopy images of the prepared devices, with white dashed boxes highlighting the channel region used in the device; **b** and **c** show the AFM and Raman characterizations, respectively. AFM measurements were done for the colored dashed lines in **a**. Raman was performed for a region pointed by a yellow circle in **a**; **d**. The distribution of the model parameter $\eta$ in the device, where the color of the block represents the exact value for $\eta$; With the classification in **d**, three typical-$\eta$ devices, Dev#5-6, Dev#1-2, and Dev#2-3, were selected, and their transfer (**e-g**), output (**h-j**), and $V_{GS}$ = 3 V drain conductance (**k-m**) curves are shown in the figure.

Accurate drain conductance is essential for circuit simulation because, as is well known, the gain of a transistor amplifier is strongly influenced by the drain conductance (whose inverse is the output impedance). Even the predicted trend would be incorrect if the contact effect is not considered correctly. As a comparison, the output and drain conductance characteristics of the $\eta$ device Dev #2-3 with the most pronounced contact effect at $V_{GS}$ = 3 V were simulated using the traditional EPR method, as shown in **Fig. 3**. Although the equivalent parasitic resistance $R_{sd}$ was scanned over a wide range of values (10 kΩ ~ 10 MΩ), the trend obtained by EPR was still inconsistent with the experimental results, significantly inferior to the Landauer-QFLPS model. As for the drain conductance, the difference between the two is even more apparent: $G_{DS}$ predicted by EPR, which monotonically decreases with $V_{DS}$, is entirely wrong, while Landauer-QFLPS can correctly reproduce the single-peak of the $G_{DS}$ curve rather than a monotonically changed pattern.

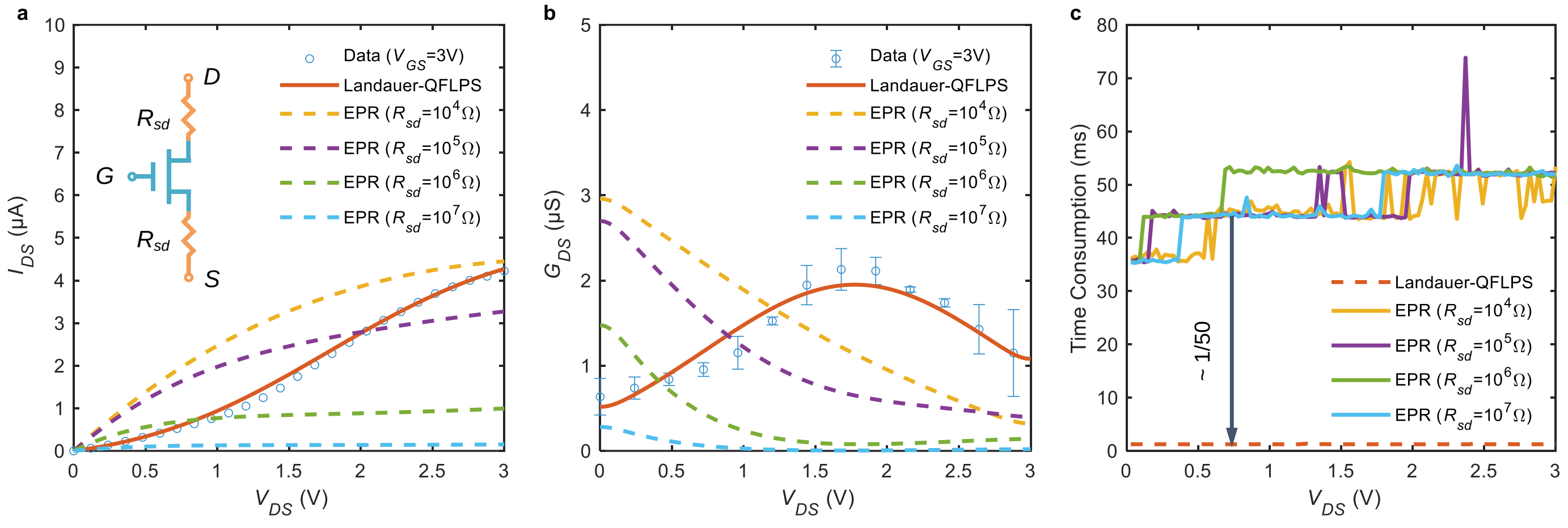

**Fig. 3 | Comparison of the Landauer-QFLPS and EPR model**s. **a**. Output characteristics; **b**. drain conductance at $V_{GS}$ = 3 V; **c**. Simulation run time. The computer processor information used for testing is Intel(R) Core (TM) i7-10700 CPU (2.90GHz).

The significant improvement in simulation quality should be attributed to the non-zero BAF. It is worth comparing the simulation results of the Landauer-QFLPS model and the EPR model with $R_{sd}$ = 10 kΩ in **Fig. 3**a to illustrate its role. The difference between them is most prominent near $V_{DS} = 0$, and as $V_{DS}$ increases, the two tend to overlap gradually. In the EPR model with such a small resistance, the BAF term is equivalently set to 0, and the electron density excited by the gate field gradually decreases as the drain potential increases due to the depletion of the gate-drain field, while the current gradually shows a concavely saturated trend due to the increased lateral drain-source field. In the Landauer-QFLPS model with a non-zero BAF, a significant contact effect represented by a large BAF near $V_{DS}$=0 strongly suppresses the output curve. When $V_{DS}$ increases to a significant level, the scattering barrier factor is almost no longer affected by $V_{DS}$, and the trend of the corrected model is consistent with that of the intrinsic model.

In addition to accuracy, it should be noted that the model's computing efficiency is significantly superior to the EPR method. This is crucial since the Landauer-QFLPS model does not require the solution of equations in order to characterize the contact effect. By comparing the CPU running time used by a single calculation obtained by averaging cumulative timings, as shown in **Fig. 3**c, it was determined that the former is roughly two orders of magnitude faster than the latter for the identical task (calculating the data in **Fig. 3**a). To guarantee a fair comparison, non-elementary calculations required by EPR and Landauer-QFLPS computations employed the identical numerical methodologies. The speed improvement is therefore solely attributable to the time saved by not needing to solve the KCL equation.

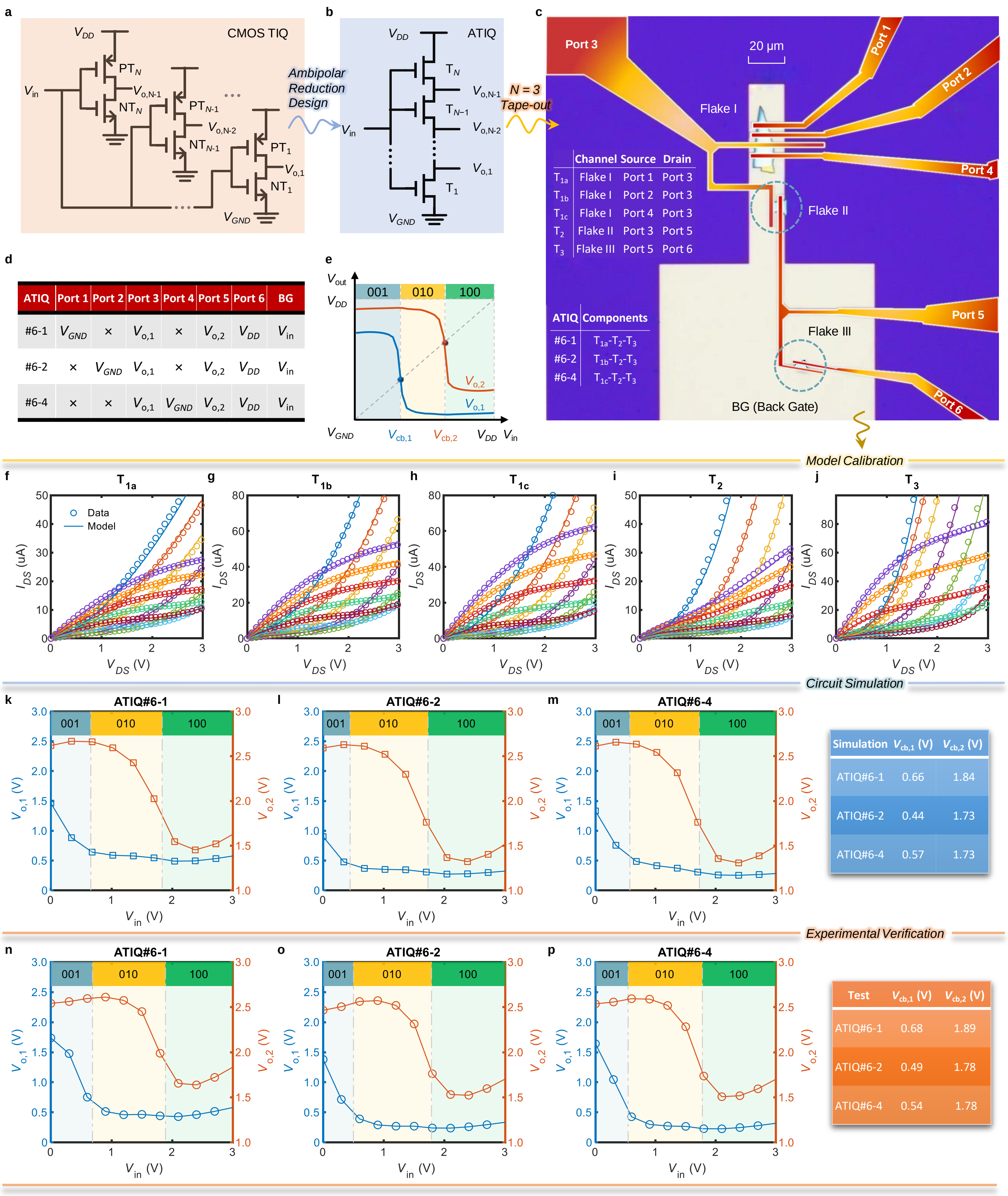

**Fig. 4 | Circuit chip design verification: ATIQ circuit**. **a**. Threshold inverter quantizer (TIQ) circuit structure based on conventional CMOS process; **b**. ATIQ circuit structure based on ambipolar-BP process; c. optical microscope image of 3-bit ATIQ chip die, where the surface contact electrode is colored with a red-yellow gradient to visually distinguish it from the back gate (pale yellow), and the light green color is the BP thin layer material; **d**. Circuit test signal code table; **e**. The ideal electrical characteristic curve of the circuit; **f-j** are the model calibration results of the on-chip BP transistor's device output characteristics; **k-m** are the predicted circuit operating characteristics based on the calibrated model, and the table on the right summarizes the predicted boundary code element voltages; **n-p** and the corresponding actual test results are shown in the table on the right.

Overall, the proposed model can satisfactorily match all the experimental measurement data obtained over the entire voltage range. In addition, this example revealed that the $\eta$ factor recovered by the model quantitatively reflects varying contact effect, emphasizing the model's benefit in considering the intrinsic channel and contact holistically.

## Tape-out verification: BP-ATIQ circuit

The previous section demonstrated that the developed Landauer-QFLPS model could achieve customized modeling of transistor performance for specific processes. This technique enables the prediction of device performance after circuit deployment, which is the basis for chip design using emerging 2DM transistors. Here, an ambipolar threshold inverter quantizer (ATIQ), designed theoretically in previous work [39], is chosen as a chip example to test the circuit simulation capability of the model. ATIQ can be used as a quantization element in a flash analog-to-digital converter (ADC), which has a compact structure and low power consumption advantages by fully utilizing the ambipolar characteristics of 2DMs [39]. A simple summary of its principal characteristics is shown in **Fig. 4**a. The conventional threshold inverter quantizer (TIQ) uses a series of inverters with different flip thresholds to quantize the analog input into digital levels [58, 59]. Because the quantization process is completed in a single step in parallel, it is the fastest ADC architecture [60]. However, the full parallel structure also rapidly increases chip area and dynamic power consumption with increasing quantized bits, thus limiting its application. ATIQ employs the physical properties of the BP-FET's ambipolar transport to strike a balance between high speed, compact size, and low power consumption. As shown in **Fig. 4**b, BP-FETs are stacked in series and the output voltage signal is still yielded in parallel, but most of BP-FETs (except the lowest and highest ones) in the series is multiplexed during operation since they can work as a p-type or n-type transistor, saving half of the devices compared with CMOS-based TIQ. The area savings are even more noteworthy. Previous study [39] has shown that the scaling properties of chip area with the number of bits have been compressed from the square law $\mathcal{O}(\mathcal{N}^2)$ to the quasi-linear law $\mathcal{O}(\mathcal{N}\ln\mathcal{N})$ by this design, where $\mathcal{N}$ represents the number of quantization levels. Additionally, dynamic power consumption is significantly reduced because all the transistors share the only one current branch drawn from the power supply $V_{DD}$. Therefore, the ATIQ circuit is a circuit design example suitable for verifying models and has practical application significance.

The chip fabricated here includes the minimum circuit configuration that can demonstrate the operations of ATIQ, which is the two-output-port structure with three BP-FETs, as shown in **Fig. 4**b. Its ideal voltage transfer curves are shown in **Fig. 4**e. Due to the different equivalent networks seen by the two ports, two different code boundary voltages $V_{\mathrm{cb},1}$ and $V_{\mathrm{cb},2}$ are successively generated, thereby realizing a three-bit one-hot-coded quantization. Without a doubt, if the large-area thin film preparation method is adopted, the quantization level that can be verified will be higher. Although the mechanical exfoliation method is more difficult in preparing the circuit, the advantage is that it can provide a more comprehensive check of the model, so this preparation method is still adopted here.

Based on the ambipolar-BP process, the core circuit unit of the 3-bit ATIQ chip was fabricated in the laboratory. The optical microscope photo of the chip is shown in **Fig. 4**c. A local back-gate (BG) was fabricated as the input port, and three flakes of the appropriate thickness (Flake I-III) were transferred to the local gate region. Among them, Flake I has a larger area, which allows the fabrication of three BP-FETs, $T_{1a}$, $T_{1b}$, and $T_{1c}$, by sharing a common drain electrode (Port 3) so that different BP-FETs can be accessed by choosing the corresponding ports for $V_{GND}$ in subsequent circuit tests. The other two smaller flakes, Flake II and Flake III are used to construct $T_2$ and $T_3$, respectively, where the drain of $T_2$ is connected to the source of $T_3$ through Port 5, the source of $T_2$ is connected to Port 3, and the drain of $T_3$ is led out through Port 6 for power supply $V_{DD}$. The high two-bit devices of the 3-bit ATIQ are fixed to $T_2$ and $T_3$. In contrast, the low-bit device is selected among $T_{1a}$, $T_{1b}$, and $T_{1c}$, which are denoted as ATIQ#6-1, ATIQ#6-2, and ATIQ#6-4, respectively. The mapping relationship among circuits, devices, and ports on the chip is summarized in the embedded table in **Fig. 4**c.

First, the I-V data of the BP-FETs on the chip were measured to calibrate the model. The bias conditions were consistent with the previous section. The comparison between the model simulations and the experimental data is shown in **Fig. 4**f-j (parameters are given in Supplementary Note 6). Again, the model can accurately reproduce the experimental results for both $T_{1a/b/c}$ and $T_3$, which show negligible contact effect, and for $T_2$, which shows significant contact effects.

Next, the three 3-bit ATIQ circuits were simulated based on the calibrated model. The test scheme is summarized in **Fig. 4**d, and is illustrated as follows. It is assigned that the input analog signal $V_{\text{in}}$ was fed from the local back gate BG, the power supply voltage $V_{DD}$ was connected to Port 6, and the output voltages $V_{\text{o},1}$and $V_{o,2}$were measured from Port 3 and Port 5, respectively. To test ATIQ#6-1, one should connected $V_{GND}$ to Port 1 while floating Port 2 and Port 4; To test ATIQ#6-2, one should connected $V_{GND}$ to Port 2 while floating Port 4 and Port 1; And, to test ATIQ#6-4, one should connected $V_{GND}$ to Port 4 while floating Port 1 and Port 2; The circuit simulation results are shown in **Fig. 4**k-m, demonstrating that the model predicts the three circuits all exhibiting the expected three-value quantization. Based on the simulations, the key circuit parameters, the code boundary voltages $V_{\text{cb},1}$ and $V_{\text{cb},2}$, were extracted, as summarized in the table beside **Fig. 4**m.

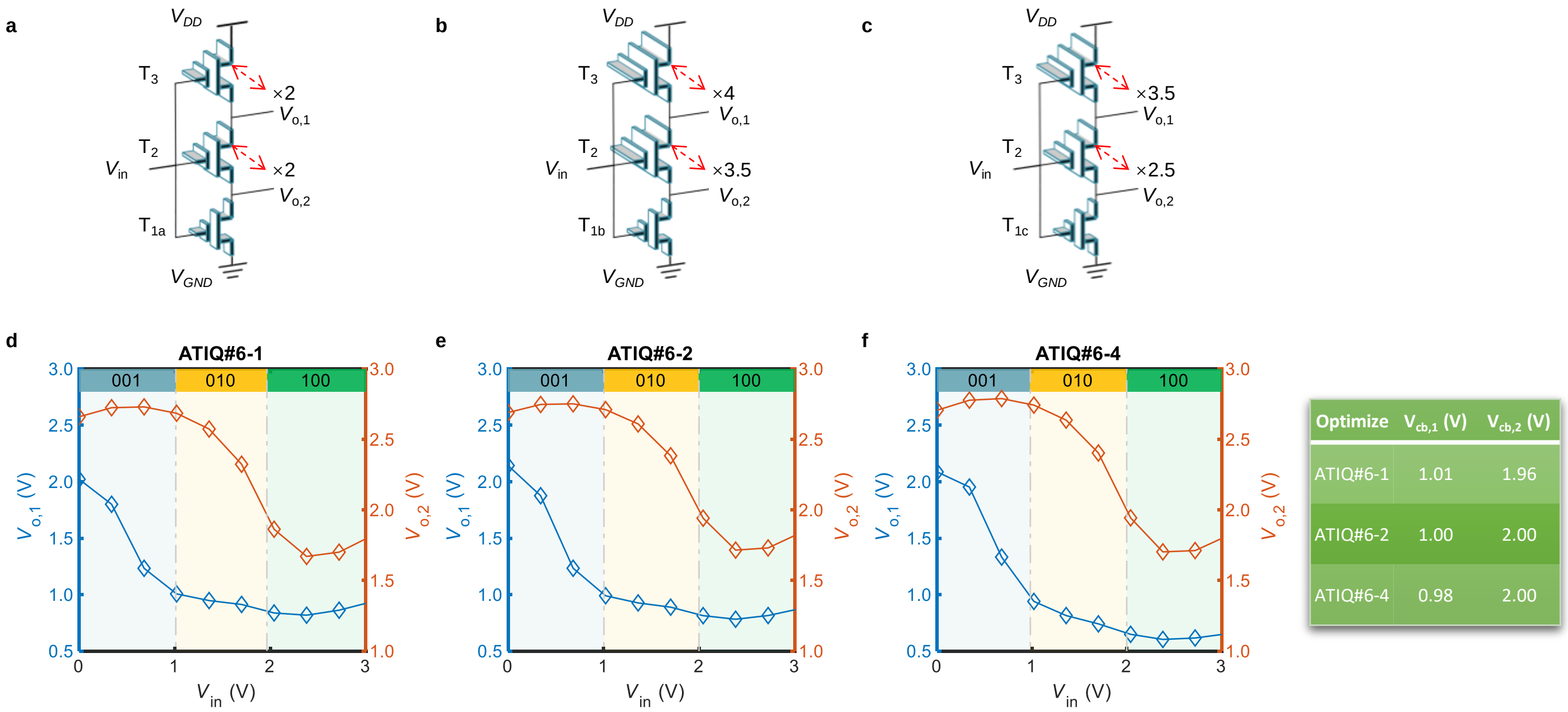

**Fig. 5 | Optimizing 3-bit ATIQ circuits based on the Landauer-QFLPS model. a-c** are schematics for the width optimizations of the three circuits ATIQ#6-1, ATIQ#6-2, and ATIQ#6-4, respectively. **d-f** are the optimized circuit operating curves. The table on the right summarizes the optimized boundary codes.

The measured data of the circuits are shown in **Fig. 4**n-p. Comparing with the simulations in **Fig. 4**k-m, it is evident that the model's simulations successfully reproduce the experimental data. The extracted code boundary voltages from the experiment show an error within 50 mV compared to the model's extracted results, thus quantitatively verifying the model's circuit simulation capability.

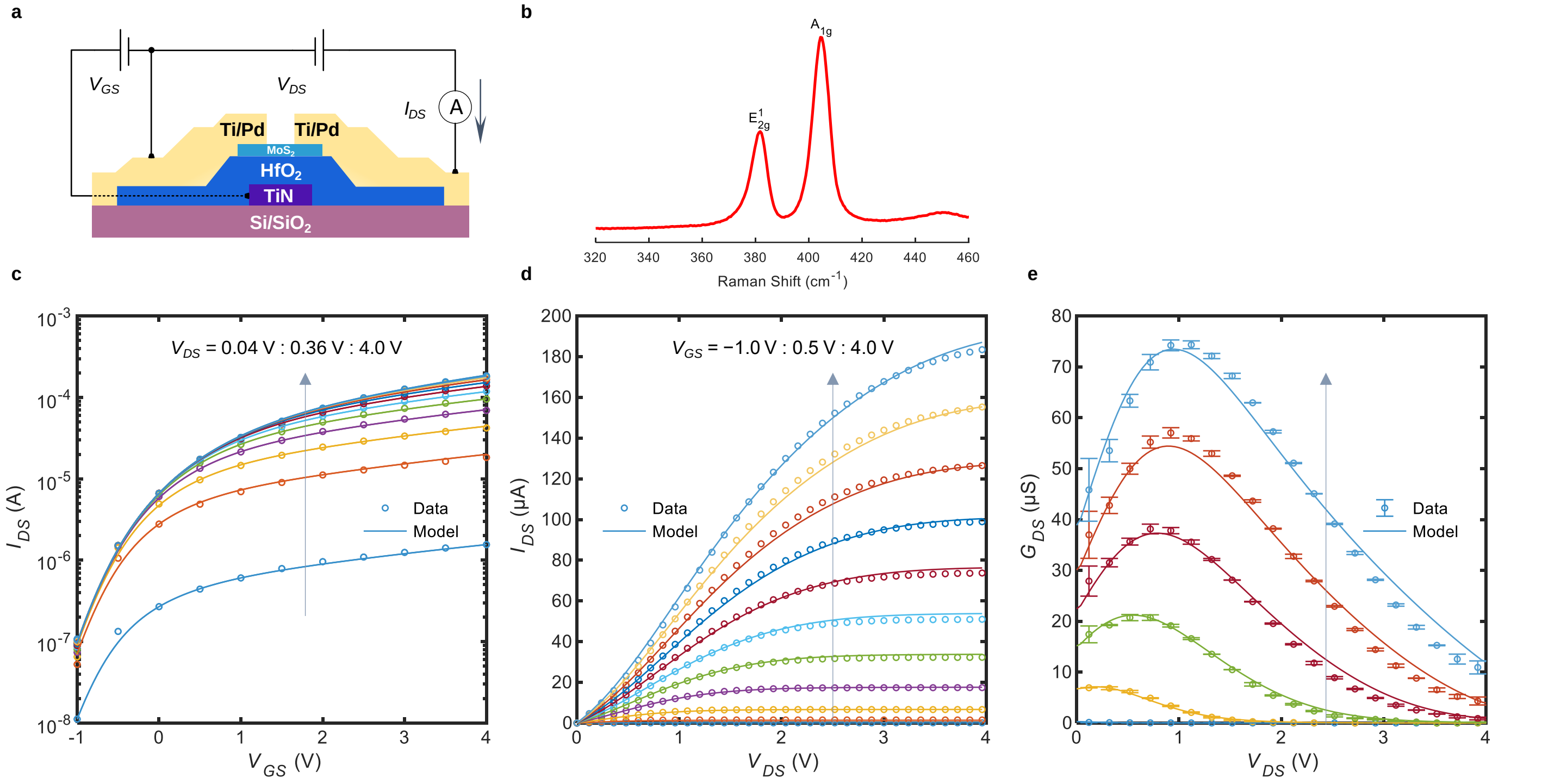

**Fig. 6 | Verification of $MoS_2$ devices**. **a**, Schematic diagram; **b**, Raman spectrum; (**c**-**e**) Comparison of simulated and measured data for the device's transfer, output, and drain conductance characteristics.

Note that the code boundary voltage of the ATIQ#6-1 configuration is higher than the other two configurations. This can be explained by the fact that the lowest transistor $T_{1a}$ of the ATIQ#6-1 has a higher impedance compared to $T_{1b}$ and $T_{1c}$ used in the other two circuits, consistent with the fact that $T_{1a}$ has the longest channel (see **Fig. 4**c).Optimizing the circuit performance is an important function of EDA software. Typically, the ideal code distribution should be uniform. That is, the overall quantization space is evenly divided. Hence, we should have $V_{\mathrm{cb},1}$ =1 V and $V_{\mathrm{cb},2}$ =2 V. However, the initial boundary codes shown above are lower than these ideal values. Although many model parameters can be adjusted to achieve the optimization goal, the most realistic optimization strategy in practical circuit design is still to adjust the channel widths. The following calculations show that this is feasible.

Note that the drain current remains unchanged under overall proportional scaling, so we can fix the scaling factor for one device and modify the channel width design of the remaining devices. Since we need to increase the boundary codes, we need to increase the conductivity of the pull-up network, so it can be expected that the correction factors for $T_2$ and $T_3$ will be greater than 1. Furthermore, due to the physical half-pitch of the process, the allowed width design is not continuous on the real axis but can only take some discrete values. Obviously, the larger the device size, the more quasi-continuous the values. However, too large dimension will impair the chip area. Now we consider an extreme case, assuming that the original device width is already twice the half-pitch, so the acceptable width-amplification factor can only be a multiple of 0.5.

Under the constraints, the device widths were redesigned using the calibrated model. The results are shown in **Fig. 5**a-c, which indicate the magnification ratio of the new widths relative to the original values. With the new designs, a sufficiently uniform code distribution was achieved (**Fig. 5**d-f), with fluctuations relative to the ideal distribution not exceeding 40 mV.

## Unipolar version: $MoS_2$-FET verification

A $MoS_2$-FET device is studied in the following to demonstrate this model for unipolar devices. $MoS_2$ is a typical n-type 2D layered semiconductor that has attracted much attention due to its suitable mobility and bandgap. Here, the $MoS_2$ transistors [61] were prepared based on back-gate technology for research purposes (**Fig. 6**a-b).

The power supply voltage was selected as 4 V. Unlike BP, the $MoS_2$ transistor completely turns off at a negatively higher gate-source voltage, so $V_{GS}$ here is only measured till −1V. The experimental data for the transfer, output, and drain conductance characteristics of the prepared $MoS_2$ transistor are shown in the circular plots in **Fig. 6**c-e. It can be seen that the output characteristics exhibit a significant contact effect, while the corresponding drain conductance curve shows a clear single-peak feature. The model extraction results (parameters given in Supplementary Note 7) show that, for unipolar devices, the model can still describe the device well and achieve first-order derivative accuracy.

## Conclusion

This article reports a theoretical model, Landauer-QFLPS model, which combines the contact current with the channel current to describe a 2DM-FET. The unique advantage of the Landauer-QFLPS model is manifested in its simulation ability to reproduce the non-monotonic drain conductance observed in the experiments. The universality and practicality of the model is examined in detail to achieve a full-stack coverage from underlying physics to top-level system applications. The Landauer-QFLPS model provides critical support for ending the long-standing situation where no standard model is available for 2DM-FETs and will actively promote 2DM transistors as a new electronic device for mature large-scale system-level applications.

## Method

The carrier density model, low-rank approximation formula, and experimental preparation processes are described below.

### Carrier density model

The electron ($n$) and hole ($p$) densities' functional relations with their QFLs ($\varepsilon_{Fn}$ and $\varepsilon_{Fp}$, respectively) are implicitly determined by the electrostatic-statistic relations (ESRs), which comprise three equations given as follows [39].

(i) Gauss's law for the gate-oxide-2DM channel system

$$q^2(n-p)/C_{ox}+\Psi+qV_{GS}=0 \tag{10}$$

where $\Psi$ is the electrostatic energy, and $C_{ox}=\epsilon_{r,ox}\epsilon_0/t_{ox}$ is the specific gate-oxide capacitance with $\epsilon_{r,ox}$ as the relative permittivity, $\epsilon_0$ as the vacuum permittivity, and $t_{ox}$ as the gate-oxide thickness.

(ii) 2D-electron density

$$n=\Phi_{\mathrm{t}}D_e\ln(1+\exp\Phi_t^{-1}(\varepsilon_{Fn}-\Psi-\Phi_n')) \tag{11}$$

(iii) 2D-hole density

$$p=\Phi_t D_h\ln\big(1+\exp\Phi_t^{-1}\big(\Psi-\Phi_p'-\varepsilon_{Fp}\big)\big) \tag{12}$$

where $\Phi_t$ is the effective thermal barrier, $D_{e(h)} = g_s g_v m^*_{e(h)} m_0/\pi\hbar^2$ is the effective density of states (DOS) for conduction (valence) band electrons (holes) with the spin valley degeneracy set as 1 for simplicity. Here, $m^*_e = 0.15$ and $m^*_h = 0.14$ are used for BP's simulations. For $MoS_2$, $m^*_e = 0.45$ and $m^*_h = 0.55$ are used. The shifted Fermi potential barrier for electrons ($\Phi'_n$) and holes ($\Phi'_p$) are respectively defined as

$$\Phi'_n = \Phi_{n0} - \Phi_{ms} - q^2 N_f/C_{ox} + q^2 N_{it,e}/C_{ox} \tag{13}$$

$$\Phi'_p = \Phi_{p0} + \Phi_{ms} + q^2 N_f/C_{ox} + q^2 N_{it,h}/C_{ox} \tag{14}$$

For BP, it is set that equilibrium electron Fermi potential $\Phi_{n0} = 0.19$ eV, equilibrium hole Fermi potential $\Phi_{p0} = 0.20$ eV, workfunction difference $\Phi_{ms} = 0.02$ eV, and fixed charged impurity $N_f = 5.03 \times 10^{12}$ $cm^{-2}$.

Efficient algorithms to solve the ESRs defined by Eqs. (10)-(12) and the integrals in Eqs. (7) and (8) have been reported by our previous numerical work [43]. Hence, there is no obstacle to planting the model into a standard circuit simulator.

## Low-rank approximation formula

An $N$-th ($N \geq 2$) order linearized Gaussian Wavelet (LGW) with $\sigma$-extension on the bounded interval $\mathcal{B} = [V_{GS,min}, V_{GS,max}]$ is constructed as follows [44]

$$H_{N,\sigma}(V_{GS}) = \int_{-\infty}^{+\infty} \frac{1}{\sqrt{2\pi}\sigma} e^{-(X-V_{GS})^2/2\sigma^2} h\{Y_i\}_{1\leq i\leq N}(X) dX \tag{15}$$

where $h\{Y_i\}_{1\leq i\leq N}(X)$ represents the linear interpolation function of the control points $\{X_i, Y_i\}_{1\leq i\leq N}$ located on interval $\mathcal{B}$, and the interpolation function value outside the interpolation interval $\mathcal{B}$ is set to 0. Although the integral form is employed by Eq. (15), the value can be evaluated analytically with the help of the error function (given in Supplementary Note 3), which possesses a standard fast algorithm by rational Chebyshev approximations [62].

### *Control points setting*

In principle, both the $\{X_i\}_{1\leq i\leq N}$ and $\{Y_i\}_{1\leq i\leq N}$ coordinates can be selected as optimization variables, but this would significantly increase the difficulty of the optimization convergence. Therefore, we evenly distribute the $\{X_i\}_{1\leq i\leq N}$ coordinates on the interpolation interval $\mathcal{B}$ and leave only the $\{Y_i\}_{1\leq i\leq N}$ coordinates as optimization variables.

### *Rank optimization*

In principle, the higher the value of $N$, the more accurate the approximation. However, due to considerations of convergence cost, we limit $N$ to be less than or equal to 3 in this case.

### *Boundary effect*

We employed the odd-extension technique in the computation here to suppress the well-known boundary effects [63].

## Fabrication process

### *BP-FET*

First, deposit a 35 nm Pd/3.5 nm Ti metal stack as a buried gate electrode on a 300 nm $SiO_2$ substrate. Then, by atomic layer deposit (ALD), a 20 nm $Al_2O_3$ dielectric layer was prepared. Next, use a mechanical exfoliation process to peel BP from the bulk material into thin layers and transfer it onto the prepared buried gate substrate using a dry transfer method. Next, a 3.5 nm Ti/35 nm drain-source electrode stack was defined using electron beam lithography. Finally, a 20 nm $Al_2O_3$ passivation layer was encapsulated to protect the BP channel.

### *$MoS_2$-FET*

First, we sputtered a 10 nm TiN electrode onto a 300 nm $SiO_2$ substrate and left a predefined pattern with a lift-off process. Next, a 15 nm $HfO_2$ dielectric layer was deposited with ALD as the gate insulator, followed by a spin-coating process to deposit a uniform layer of $MoS_2$ nanosheet [61]. Heat the sample in an $N_2$ atmosphere for 1 hour at 300 ℃. Finally, we used electron beam evaporation to deposit source/drain Ti/Pd electrodes.

## Acknowledgments

This work is supported by the National Key R&D Program (2022YFB3204100), National Natural Science Foundation (U20A20168, 51861145202, 61874065, 62022047, 62274101, 61974049) of China, JCCDFSIT 2022CDF003, and the Foundation of State Key Laboratory of High-efficiency Utilization of Coal and Green Chemical Engineering (Grant No. 2022-K81).

# Landauer-QFLPS model for mixed Schottky-Ohmic Contact two-dimensional transistors
# Supplementary Note

Zhao-Yi Yan,[1,2,†] Zhan Hou,[1,2,†] Kan-Hao Xue,[3,4,*] Tian Lu,[1,2] Ruiting Zhao,[1,2] Junying Xue,[5] Fan Wu,[1,2] Minghao Shao,[1,2] Jianlan Yan,[1,2] Anzhi Yan,[1,2] Zhenze Wang,[1,2] Penghui Shen,[1,2] Mingyue Zhao,[1,2] Xiangshui Miao,[3,4] Zhaoyang Lin,[5] Houfang Liu,[1,2,*] He Tian,[1,2,*] Yi Yang,[1,2,*] Tian-Ling Ren[1,2,*]

[1] *School of Integrated Circuits, Tsinghua University, Beijing 100084, China*

[2] *Beijing National Research Center for Information Science and Technology (BNRist), Tsinghua University, Beijing 100084, China*

[3] *School of Integrated Circuits, School of Optical and Electronic Information, Huazhong University of Science and Technology, Wuhan 430074, China*

[4] *Hubei Yangtze Memory Laboratories, Wuhan 430205, China*

[5] *Department of Chemistry, Tsinghua University, Beijing 100084, China*

***Corresponding Author**, E-mail: RenTL@tsinghua.edu.cn (T.-L. Ren), yiyang@tsinghua.edu.cn (Y. Yang), tianhe88@tsinghua.edu.cn (H. Tian), houfangliu@tsinghua.edu.cn (H. Liu), xkh@hust.edu.cn (K.-H. Xue)

[†]These authors contributed equally.

Formula derivations, TEM & EDS characterizations for phosphorus oxide, extracted model parameters, and completed I-V data are given in this note.

## Supplementary Note 1 | Landauer-QFLPS formula derivation: electrons flow

Based on the Landauer formula, the electron flow injected from the source junction is written as

$$I_{es} = W\frac{q}{\pi\hbar}\int_{-\infty}^{+\infty}\Gamma_{es}(\varepsilon)M_{es}(\varepsilon)[f(\varepsilon,\varepsilon_{Fs}) - f(\varepsilon,\varepsilon_{Fni})]d\varepsilon \qquad (S1)$$

where $\varepsilon_{Fni}$ and $\varepsilon_{Fs}$ label the Fermi levels of the intrinsic channel point and the source electrode, respectively. The Fermi-Dirac distribution function is defined as

$$f(\varepsilon,\varepsilon_F) = \frac{1}{1+\exp\left(\frac{\varepsilon-\varepsilon_F}{kT}\right)} \qquad (S2)$$

where $k$ and $T$ denotes the Boltzmann constant and the temperature, respectively. The modeling for the transmission function $\Gamma_{es}(\varepsilon)$ and the effective density of mode (DOM) function $M_{es}(\varepsilon)$ are given as follows.

A. transmission function $\Gamma_{es}(\varepsilon)$

The transmission rate function $\Gamma_{es}(\varepsilon)$ considers two kinds of transport mechanism: (i) quantum propagation rate that happens on the global energy scale, and (ii) thermal emission rate that occurs only when the energy is higher than the conduction band minimum $\varepsilon > E_{cs}$;

For quantum propagation rate, WKB approximation ($\varepsilon < E_{cs}$) gives that

$$\Gamma_{es}(\varepsilon) = \exp(-2\gamma) \qquad (S3)$$

where the $\gamma$ factor is calculated as

$$\gamma(\varepsilon) = \pm\int_0^a \frac{1}{\hbar}\sqrt{|2m_e^* m_0(E_c(x)-\varepsilon)|}\,dx \qquad (S4)$$

where $E_c(x)$ represents the energy profile function and $a$ denotes the endpoint of the propagation path at the energy level $\varepsilon$ so that $E_c(a) = \varepsilon$ and $E_c(0) = E_{cs}$. Energy band profile $\varepsilon_c(x)$ can be determined with Poisson's equation as a parabolic solution as

$$E_c(x) = \varepsilon + \frac{q\rho_s}{\epsilon_s}(x-a)^2 \qquad (S5)$$

Substituting the expression into $\gamma(\varepsilon)$, one obtains that

$$\gamma(\varepsilon) = \frac{1}{\hbar}\frac{E_{cs}-\varepsilon}{\sqrt{2q\rho_s/m_e^* m_0\epsilon_s}} \qquad (S6)$$

For $\varepsilon > E_{cs}$, the transmission rate is enhanced by the thermal emission mechanism as

$$\Gamma_{es}(\varepsilon) = \exp\big(-2\gamma + (\varepsilon - E_{cs})/\Phi_{t,es}\big) \qquad (S7)$$

where $\Phi_{t,es}$ is the thermal emission barrier.

B. density of modes (DOM) function $M_{es}(\varepsilon)$

The DOM function is defined as

$$M_{es}(\varepsilon) = \frac{g_v}{\pi h}\sqrt{2m_{es}^* m_0 K_{es}(\varepsilon)} \qquad (S8)$$

where $g_v$ is the valley degeneracy, $m_{es}^*$ is the relative effective mass for the electrons in the source metal, and $K_{es}(\varepsilon)$ is the collective kinectic energy of the source electron defined as

$$K_{es}(\varepsilon) = E_{b,es} \exp\left(-\frac{\varepsilon - \varepsilon_{Fs}}{kT_s}\right) \exp\left(\frac{qV_{DS}}{\Phi_{a,es}}\right) \tag{S9}$$

where $E_{b,es}$ is the elemental thermal kinectic energy for electron in the source metal, $T_s$ is the effectively local temperature for the source junction, and $\Phi_{a,es}$ is the acceleration barrier for the electrons injected from the source junction.

C. Formulization of $I_{es}$

Given the definitions for $\Gamma_{es}(\varepsilon)$ (Eq. $(S3)$ and $(S7)$) and $M_{es}(\varepsilon)$ (Eq. $(S8)$), one can introduce an index function $A(\varepsilon)$ to simply the calculation so that

$$\Gamma_{es}(\varepsilon) M_{es}(\varepsilon) = e^{A(\varepsilon)} [\mathrm{m}^{-1}] \tag{S10}$$

which leads to the explicit expressions for $A(\varepsilon)$ as

$$A(\varepsilon) = \begin{cases} -2\gamma + \dfrac{\varepsilon - E_{cs}}{\Phi_{t,es}} - \dfrac{\varepsilon - \varepsilon_{Fs}}{2kT_s} + \ln M_{es}(\varepsilon_{Fs}) & \varepsilon \geq E_{cs}, \\ -2\gamma - \dfrac{\varepsilon - \varepsilon_{Fs}}{2kT_s} + \ln M_{es}(\varepsilon_{Fs}) & \varepsilon < E_{cs}, \end{cases} \tag{S11}$$

With the aid of $A(\varepsilon)$, one can go further to simplify the $I_{es}$ in Eq. $(S1)$ as

$$I_{es} = W \frac{q}{\pi\hbar} \int_{-\infty}^{+\infty} e^{A(\varepsilon)} [f(\varepsilon, \varepsilon_{Fs}) - f(\varepsilon, \varepsilon_{Fni})] d\varepsilon \tag{S12}$$

Re-write the Fermi window in the square bracket as the integral of the derivative as

$$f(\varepsilon, \varepsilon_{Fs}) - f(\varepsilon, \varepsilon_{Fni}) = \int_{\varepsilon_{Fni}}^{\varepsilon_{Fs}} \frac{\partial}{\partial \varepsilon_F} f(\varepsilon, \varepsilon_F) \, d\varepsilon_F \tag{S13}$$

Since the 2D electron density $n$ is proportional to the integral of the $f$ as

$$n(E_c, \varepsilon_F) = \int_{E_c}^{+\infty} D_e f(\varepsilon, \varepsilon_F) \, d\varepsilon \tag{S14}$$

where $D_e = g_s g_v m_e^* m_0 / \pi\hbar^2$ is the density of states for conduction band electrons in the channel semiconductor with the spin valley degeneracy set as 1, distribution function $f$ can be expressed by the partial derivative of $n$ as

$$f(\varepsilon, \varepsilon_F) = \frac{1}{D_e} \frac{\partial}{\partial \varepsilon_F} n(\varepsilon, \varepsilon_F) \tag{S15}$$

and an useful identity holds as

$$\frac{\partial}{\partial \varepsilon_F} f(\varepsilon, \varepsilon_F) = -\frac{\partial}{\partial \varepsilon} f(\varepsilon, \varepsilon_F) \tag{S16}$$

Next, substituting Eq. $(S13)$ into Eq. $(S12)$ and interchange the order of integrals, one obtains

$$I_{es} = W \frac{q}{\pi\hbar} \frac{1}{D_e} \int_{\varepsilon_{Fni}}^{\varepsilon_{Fs}} \left\{ \int_{-\infty}^{+\infty} e^{A(\varepsilon)} \frac{\partial}{\partial \varepsilon_F} f(\varepsilon, \varepsilon_F) d\varepsilon \right\} d\varepsilon_F \tag{S17}$$

Using properties $(S15)$ and $(S16)$, one has

$$I_{es} = W \frac{q}{\pi\hbar} \frac{1}{D_e} \int_{\varepsilon_{Fni}}^{\varepsilon_{Fs}} \left\{ \int_{-\infty}^{+\infty} e^{A(\varepsilon)} \frac{\partial^2}{\partial \varepsilon^2} n(\varepsilon, \varepsilon_F) d\varepsilon \right\} d\varepsilon_F \tag{S18}$$

Integrating by parts for the inner bracket integral yields

$$I_{es} = W \frac{q}{\pi\hbar} \frac{1}{D_e} \int_{\varepsilon_{Fni}}^{\varepsilon_{Fs}} \left\{ \int_{-\infty}^{+\infty} n(\varepsilon, \varepsilon_F) \, [A''(\varepsilon) + A'(\varepsilon)^2] e^{A(\varepsilon)} d\varepsilon \right\} d\varepsilon_F \qquad (S19)$$

Since a leap of the derivative arises for $A(\varepsilon)$ (c.f. Eq. $(S11)$), $A''(\varepsilon)$ yields a Dirac delta function as

$$A''(\varepsilon) = \frac{1}{\Phi_{t,es}} \delta(\varepsilon - E_{cs}) \qquad (S20)$$

Hence, by neglecting the first-order derivative of $A(\varepsilon)$, one arrives at

$$I_{es} = W \frac{q}{\pi\hbar} \frac{1}{D_e} \int_{\varepsilon_{Fni}}^{\varepsilon_{Fs}} n(E_{cs}, \varepsilon_F) \frac{1}{\Phi_{t,es}} e^{A(E_{cs})} \, d\varepsilon_F \qquad (S21)$$

With the definition of $A(\varepsilon)$ and noting that $E_{cs} - \varepsilon_{Fs} = \Phi_{sb,es}$, one further obtains

$$I_{es} = W \frac{q}{\pi\hbar} \frac{1}{D_e} \frac{1}{\Phi_{t,es}} M_{es}(\varepsilon_{Fs}) \exp\left(-\frac{\Phi_{sb,es}}{2kT_s}\right) \int_{\varepsilon_{Fni}}^{\varepsilon_{Fs}} n(E_{cs}, \varepsilon_F) \, d\varepsilon_F \qquad (S22)$$

By gathering the coefficients, it can be written as

$$I_{es} = e^{-\eta_{es} + \frac{qV_{DS}}{\Phi_{a,es}}} \frac{W}{L} \int_{\varepsilon_{Fni}}^{\varepsilon_{Fs}} \mu_n n \, d\varepsilon_F \qquad (S23)$$

where approximation $n(E_{cs}, \varepsilon_F) \approx n(E_c, \varepsilon_F)$ has been made, and the index $\eta_{es}$ is defined as

$$\eta_{es} = \ln \left[ \frac{m_e^* m_0 \, \mu_n / q}{L \frac{\sqrt{2 m_{es}^* m_0 E_{b,es}}}{8\pi^2 \Phi_{t,es}} \exp\left(-\frac{\Phi_{sb,es}}{2kT_s}\right)} \right] \qquad (S24)$$

Combining with the QEA-model for $I_e$ as

$$I_e = \frac{W}{L} \int_{\varepsilon_{Fd}}^{\varepsilon_{Fni}} \mu_n n \, d\varepsilon_F = \frac{W}{L} \int_{\varepsilon_{Fd}}^{\varepsilon_{Fs}} \mu_n n \, d\varepsilon_F - \int_{\varepsilon_{Fni}}^{\varepsilon_{Fs}} \mu_n n \, d\varepsilon_F \qquad (S25)$$

one can obtain the final result for $I_e$ by substituting the second integral in Eq. $(S25)$ with Eq. $(S23)$.

## Supplementary Note 2 | Landauer-QFLPS formula derivation: holes flow

Based on the Landau formula, the hole flow injected from drain junction is written as

$$I_{hd} = W\frac{q}{\pi\hbar}\int_{-\infty}^{+\infty}\Gamma_{hd}(\varepsilon)M_{hd}(\varepsilon)\left[f\left(\varepsilon,\varepsilon_{Fpi}\right)-f(\varepsilon,\varepsilon_{Fd})\right]d\varepsilon \tag{S26}$$

where $\varepsilon_{Fpi}$ and $\varepsilon_{Fd}$ label the Fermi levels of the intrinsic channel point and the drain electrode, respectively. The Fermi-Dirac distribution function is defined as

$$f(\varepsilon,\varepsilon_F) = \frac{1}{1+\exp\left(\frac{\varepsilon-\varepsilon_F}{kT}\right)} \tag{S27}$$

where $k$ and $T$ denotes the Boltzmann constant and the temperature, respectively. The modeling for the transmission function $\Gamma_{hd}(\varepsilon)$ and the density of mode (DOM) function $M_{hd}(\varepsilon)$ are given as follows.

A. transmission function $\Gamma_{hd}(\varepsilon)$

The transmission rate function $\Gamma_{hd}(\varepsilon)$ considers two kinds of transport mechanism: (i) quantum propagation rate that happens on the global energy scale, and (ii) thermal emission rate that occurs only when the energy is lower than the valence band maximum $\varepsilon < E_{vd}$;

For quantum propagation rate, WKB approximation ($\varepsilon > E_{vd}$) gives that

$$\Gamma_{hd}(\varepsilon) = \exp(-2\gamma) \tag{S28}$$

where the $\gamma$ factor is calculated as

$$\gamma(\varepsilon) = \pm\int_0^a \frac{1}{\hbar}\sqrt{\left|2m_h^* m_0(E_v(x)-\varepsilon)\right|}\,dx \tag{S29}$$

where $E_v(x)$ represents the valence band energy profile function and $a$ denotes the endpoint of the propagation path at the energy level $\varepsilon$ so that $E_v(a) = \varepsilon$ and $E_v(0) = E_{vd}$. Energy band profile $E_v(x)$ is described with a parabolic function as

$$E_v(x) = \varepsilon - \frac{q\rho_s}{\epsilon_s}(x-a)^2 \tag{S30}$$

Substituting the expression into $\gamma(\varepsilon)$, one obtains that

$$\gamma(\varepsilon) = \frac{1}{\hbar}\frac{-E_{vd}+\varepsilon}{\sqrt{2q\rho_s/m_h^* m_0\epsilon_s}} \tag{S31}$$

For $\varepsilon < E_{vd}$, the transmission rate is enhanced by the thermal emission mechanism as

$$\Gamma_{hd}(\varepsilon) = \exp\left(-2\gamma + (-\varepsilon + E_{vd})/\Phi_{t,hd}\right) \tag{S32}$$

where $\Phi_{t,hd}$ is the thermal emission barrier.

B. density of modes (DOM) function $M_{hd}(\varepsilon)$

The DOM function is determined by the conservation law as

$$M_{hd}(\varepsilon) = \frac{g_v}{\pi h}\sqrt{2m_{hd}^* m_0 K_{hd}(\varepsilon)} \tag{S33}$$

where $g_v$ is the valley degeneracy, $m_{hd}^*$ is the relative effective mass for the electrons in the source metal, and $K_{hd}(\varepsilon)$ is the collective kinectic energy of the source electron defined as

$$K_{hd}(\varepsilon) = E_{b,hd}\exp\left(\frac{\varepsilon-\varepsilon_{Fd}}{kT_d}\right)\exp\left(\frac{qV_{DS}}{\Phi_{a,hd}}\right) \tag{S34}$$

where $E_{b,hd}$ is the elemental thermal kinetic energy for hole in the drain metal, $T_d$ is the effectively local temperature for the drain metal, and $\Phi_{a,hd}$ is the acceleration barrier for the holes injected from the drain junction.

C. Formulization of $I_{hd}$

Given the definitions for $\Gamma_{hd}(\varepsilon)$ and $M_{hd}(\varepsilon)$, one can introduce an index function $A(\varepsilon)$ to simply the calculation so that

$$\Gamma_{hd}(\varepsilon)M_{hd}(\varepsilon) = e^{A(\varepsilon)}\ [\mathrm{m}^{-1}] \tag{S35}$$

which leads to the explicit expressions for $A(\varepsilon)$ as

$$A(\varepsilon) = \begin{cases} -2\gamma + \dfrac{-\varepsilon + E_{vd}}{\Phi_{t,hd}} + \dfrac{\varepsilon - \varepsilon_{Fd}}{2kT_d} + \ln M_{hd}(\varepsilon_{Fd}) & -\varepsilon \geq -E_{vd} \\ -2\gamma + \dfrac{\varepsilon - \varepsilon_{Fd}}{2kT_d} + \ln M_{hd}(\varepsilon_{Fd}) & -\varepsilon < -E_{vd} \end{cases} \tag{S36}$$

With the aid of $A(\varepsilon)$, one can go further to simplify the $I_{hd}$ in Eq. $(S1)$ as

$$I_{hd} = W\frac{q}{\pi\hbar}\int_{-\infty}^{+\infty} e^{A(\varepsilon)}\left[f\left(\varepsilon, \varepsilon_{Fpi}\right) - f(\varepsilon, \varepsilon_{Fd})\right]d\varepsilon \tag{S37}$$

Re-write the Fermi window in the square bracket as the integral of the derivative as

$$f\left(\varepsilon, \varepsilon_{Fpi}\right) - f(\varepsilon, \varepsilon_{Fd}) = \int_{\varepsilon_{Fd}}^{\varepsilon_{Fpi}} \frac{\partial}{\partial \varepsilon_F} f(\varepsilon, \varepsilon_F)\, d\varepsilon_F \tag{S38}$$

Since the 2D hole density $p$ formally is proportional to the integral of the $f$ as

$$p(E_v, \varepsilon_F) = \int_{-\infty}^{E_v} D_h f(-\varepsilon, -\varepsilon_F)\, d\varepsilon \tag{S39}$$

where $D_h = g_s g_v m_h^* m_0/\pi\hbar^2$ is the density of states for valence band holes in the channel semiconductor with the spin valley degeneracy set as 1, distribution function $f$ can be expressed by the partial derivative of $p$ as

$$f(-\varepsilon, -\varepsilon_F) = \frac{1}{D_h}\frac{\partial}{\partial \varepsilon} p(\varepsilon, \varepsilon_F) \tag{S40}$$

and an useful identity holds as

$$\frac{\partial}{\partial \varepsilon_F} f(\varepsilon, \varepsilon_F) = -\frac{\partial}{\partial \varepsilon_F} f(-\varepsilon, -\varepsilon_F) = \frac{\partial}{\partial \varepsilon} f(-\varepsilon, -\varepsilon_F) \tag{S41}$$

Next, similarly, one obtains

$$I_{hd} = W\frac{q}{\pi\hbar}\int_{\varepsilon_{Fd}}^{\varepsilon_{Fpi}} \left\{\int_{-\infty}^{+\infty} e^{A(\varepsilon)} \frac{\partial}{\partial \varepsilon_F} f(\varepsilon, \varepsilon_F) d\varepsilon\right\} d\varepsilon_F \tag{S42}$$

followed by

$$I_{hd} = W\frac{q}{\pi\hbar}\frac{1}{D_h}\int_{\varepsilon_{Fd}}^{\varepsilon_{Fpi}} \left\{\int_{-\infty}^{+\infty} e^{A(\varepsilon)} \frac{\partial^2}{\partial \varepsilon^2} p(\varepsilon, \varepsilon_F) d\varepsilon\right\} d\varepsilon_F \tag{S43}$$

Integrating by parts for the inner bracket integral yields

$$I_{hd} = W\frac{q}{\pi\hbar}\frac{1}{D_h}\int_{\varepsilon_{Fd}}^{\varepsilon_{Fpi}} \left\{\int_{-\infty}^{+\infty} p(\varepsilon, \varepsilon_F)\,[A''(\varepsilon) + A'(\varepsilon)^2] e^{A(\varepsilon)} d\varepsilon\right\} d\varepsilon_F \tag{S44}$$

Since a jump of the derivative arises for $A(\varepsilon)$ in this section, $A''(\varepsilon)$ yields a Dirac delta function as

$$A''(\varepsilon) = \frac{1}{\Phi_{t,hd}} \delta(-\varepsilon + E_{vd}) = \frac{1}{\Phi_{t,hd}} \delta(\varepsilon - E_{vd}) \tag{S45}$$

Hence, by neglecting the first-order derivative of $A(\varepsilon)$, one arrives at

$$I_{hd} = W \frac{q}{\pi\hbar} \frac{1}{D_h} \int_{\varepsilon_{Fd}}^{\varepsilon_{Fpi}} p(E_{vd}, \varepsilon_F) \frac{1}{\Phi_{t,hd}} e^{A(E_{vd})} \, d\varepsilon_F \tag{S46}$$

With the definition of $A(\varepsilon)$ and noting that $E_{vd} - \varepsilon_{Fd} = -\Phi_{sb,hd}$, one further obtains

$$I_{hd} = W \frac{q}{\pi\hbar} \frac{1}{D_h} \frac{1}{\Phi_{t,hd}} M_{hd}(\varepsilon_{Fs}) \exp\left(-\frac{\Phi_{sb,hd}}{2kT_d}\right) \int_{\varepsilon_{Fd}}^{\varepsilon_{Fpi}} p(E_{vd}, \varepsilon_F) \, d\varepsilon_F \tag{S47}$$

By gathering the coefficients, it can be written as

$$I_{hd} = e^{-\eta_{hd} + \frac{qV_{DS}}{\Phi_{a,hd}}} \frac{W}{L} \int_{\varepsilon_{Fd}}^{\varepsilon_{Fpi}} \mu_p p \, d\varepsilon_F \tag{S48}$$

where

$$\eta_{hd} = \ln\left[\frac{m_h^* m_0 \, \mu_p / q}{L \frac{\sqrt{2m_{hd}^* m_0 E_{b,hd}}}{8\pi^2 \Phi_{t,hd}} \exp\left(-\frac{\Phi_{sb,hd}}{2kT_d}\right)}\right] \tag{S49}$$

Combining with the QEA-model for $I_h$ as

$$I_h = \frac{W}{L} \int_{\varepsilon_{Fpi}}^{\varepsilon_{Fs}} \mu_p p \, d\varepsilon_F = \frac{W}{L} \int_{\varepsilon_{Fd}}^{\varepsilon_{Fs}} \mu_p p \, d\varepsilon_F - \int_{\varepsilon_{Fd}}^{\varepsilon_{Fpi}} \mu_p p \, d\varepsilon_F \tag{S50}$$

one can obtain the final result for $I_h$ by substituting the second integral above with Eq. ($S48$).

## Supplementary Note 3 | Linearized-Gaussian control function

Define the control set for model variables $\alpha \in \{\mu_n, N_{trp,e}, \mu_p, N_{trp,h}, \Phi_t, \varphi_a, \Phi_a\}$, where $\varphi_a \coloneqq \eta \cdot \Phi_a$, on gate-source voltage interval $[-3,3]$

$$Y\alpha = \{(v_i, y_i) | 1 \le i \le N\} \tag{S51}$$

where $v_i$ is the $N$-division points of $[-3,3]$. $Y\alpha$ set is augmented by bilateral inversion operation to obtain its closure $Y\bar{\alpha}$ to suppress the boundary effect. The $Y\alpha$ set for all devices are listed in Table SI.

The $N$-points control set generates the linearized-Gaussian wavelet control function as

$$H_{N,\sigma}(V_{GS})(v) = \sum_{y_i \in Y\bar{\alpha}} \frac{1}{2} z_i(v)[\mathrm{erf}(u_{i+1}) - \mathrm{erf}(u_i)] + \frac{\xi_i \sigma}{\sqrt{2\pi}}\left[\exp\left(-u_i^2\right) - \exp\left(-u_{i+1}^2\right)\right] \tag{S52}$$

where $\mathrm{erf}(\cdot)$ is the error function with $\xi_i = (y_{i+1} - y_i)/(v_{i+1} - v_i)$, $z_i = y_i + \xi_i(v - v_i)$, and $u_i = (v_i - v)/2\sigma$. The control function is linear with the control set, and is a quasi-affine transform of the control set. Only small control sets are required ($N = 2, 3$), hence model efficiency is kept.

## Supplementary Note 4 | TEM and EDS characterizations for the BP-FET devices

TEM analysis is performed for the cross-section of the BP-FET devices used in main text, as indicated in Fig. S1.

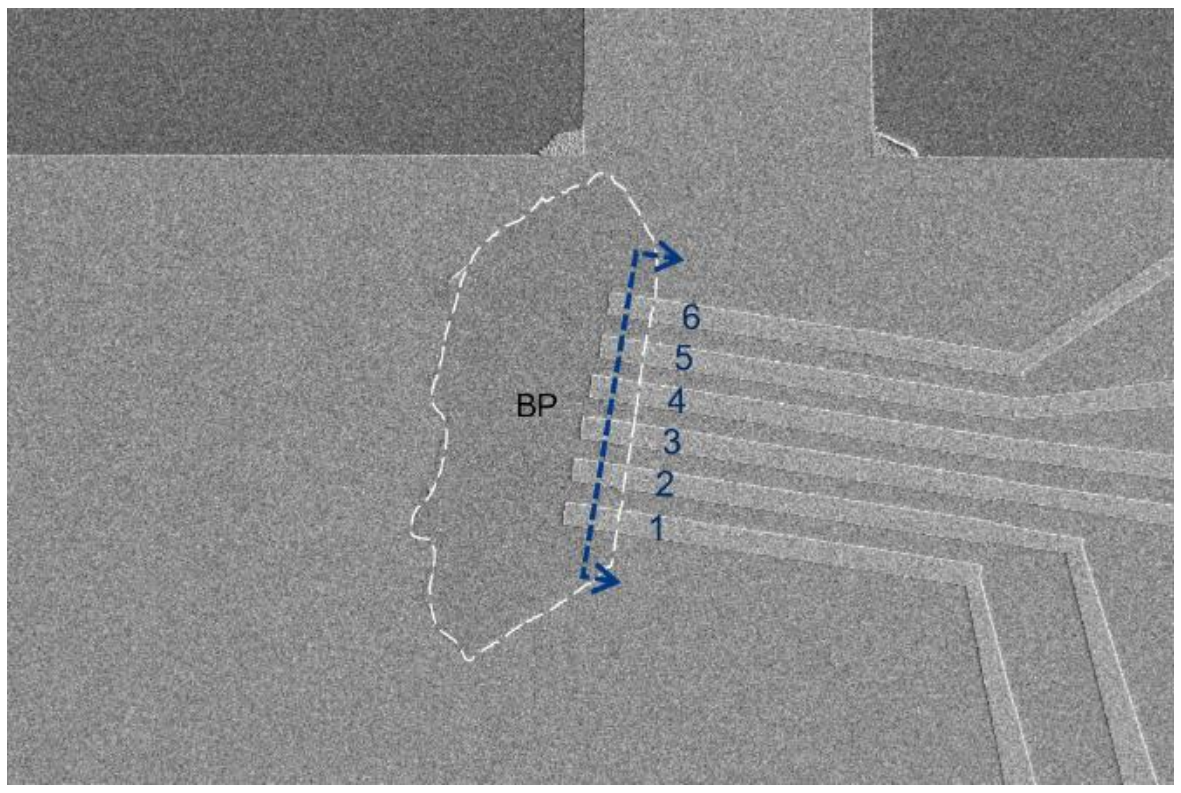

Fig. S1 The cut-line (blue dashed line) for TEM analysis

TEM graphs for the defined cross-section are shown in Fig. S2. The gray scale graphs in Fig. S2 exhibit the vertical structures of Pd/Ti/BP/$Al_2O_3$/Pd stacked layers under six electrodes labelled in Fig. S1, respectively. The 9-nm BP (as measured by AFM in the main text) contains 2-nm around $PO_x$ (varying from 1.3 nm to 2.4 nm as shown in Fig. S2).

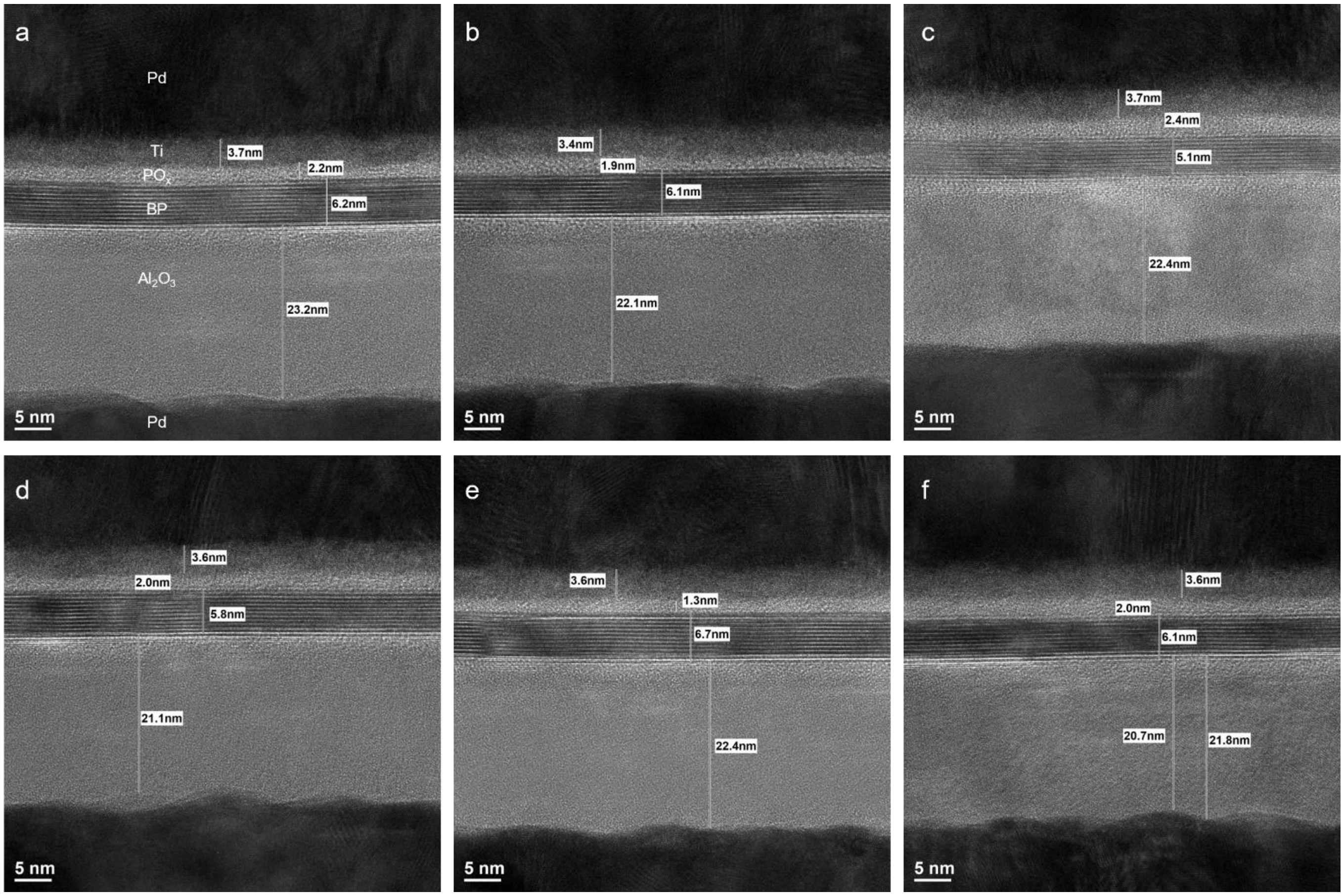

Fig. S2 Cross-section view of the BP-FET devices at 5-nm scale. Subplots **a**-**f** correspond to the electrodes 1-6.

To confirm the element species, we performed the EDS mapping for the stacked layers, as shown in Fig. S3 and Fig. S4. It shows that oxygen exists at the interfaces of BP with Ti- and Pd- layers.

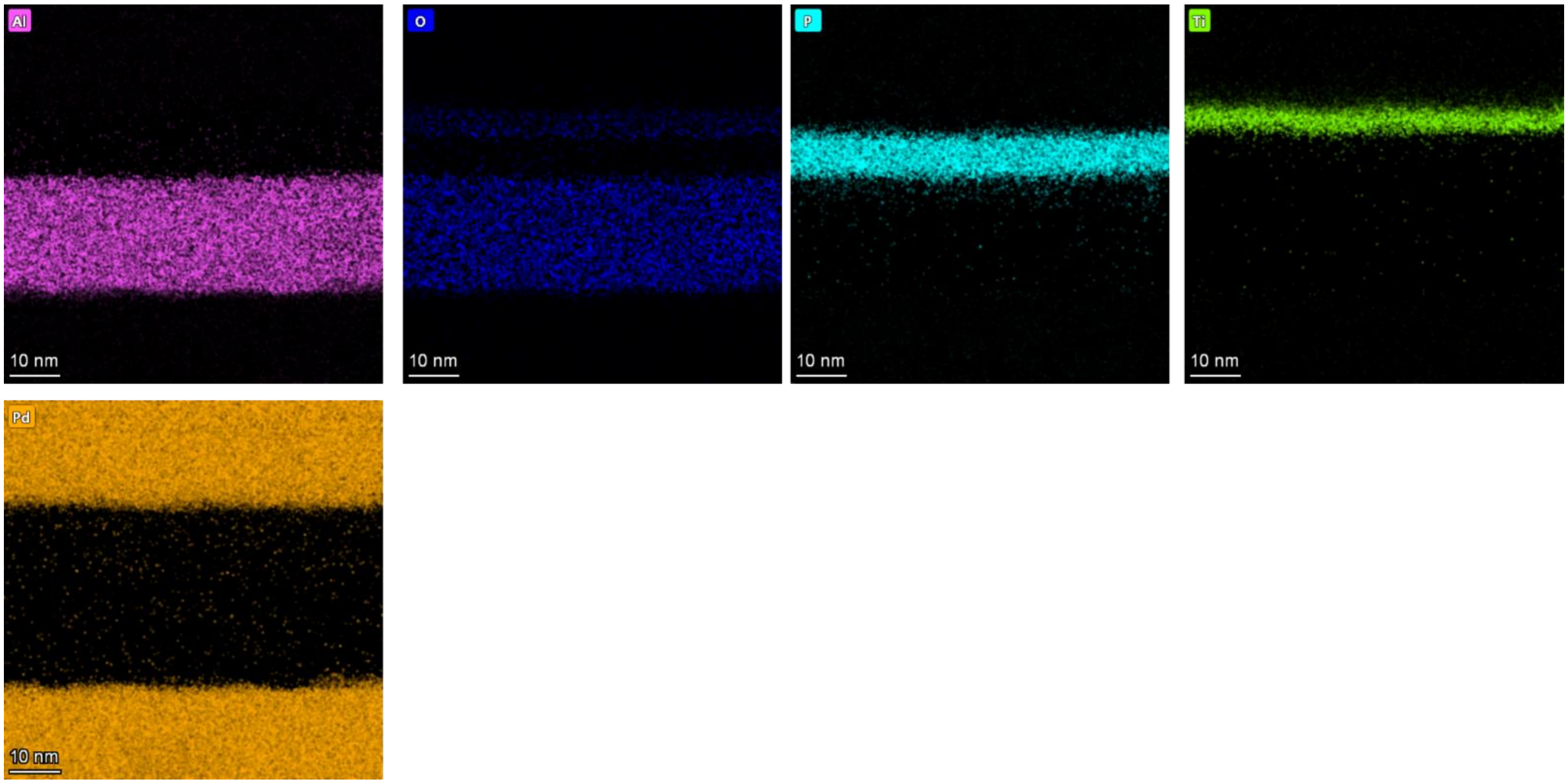

Fig. S3 EDS mapping images

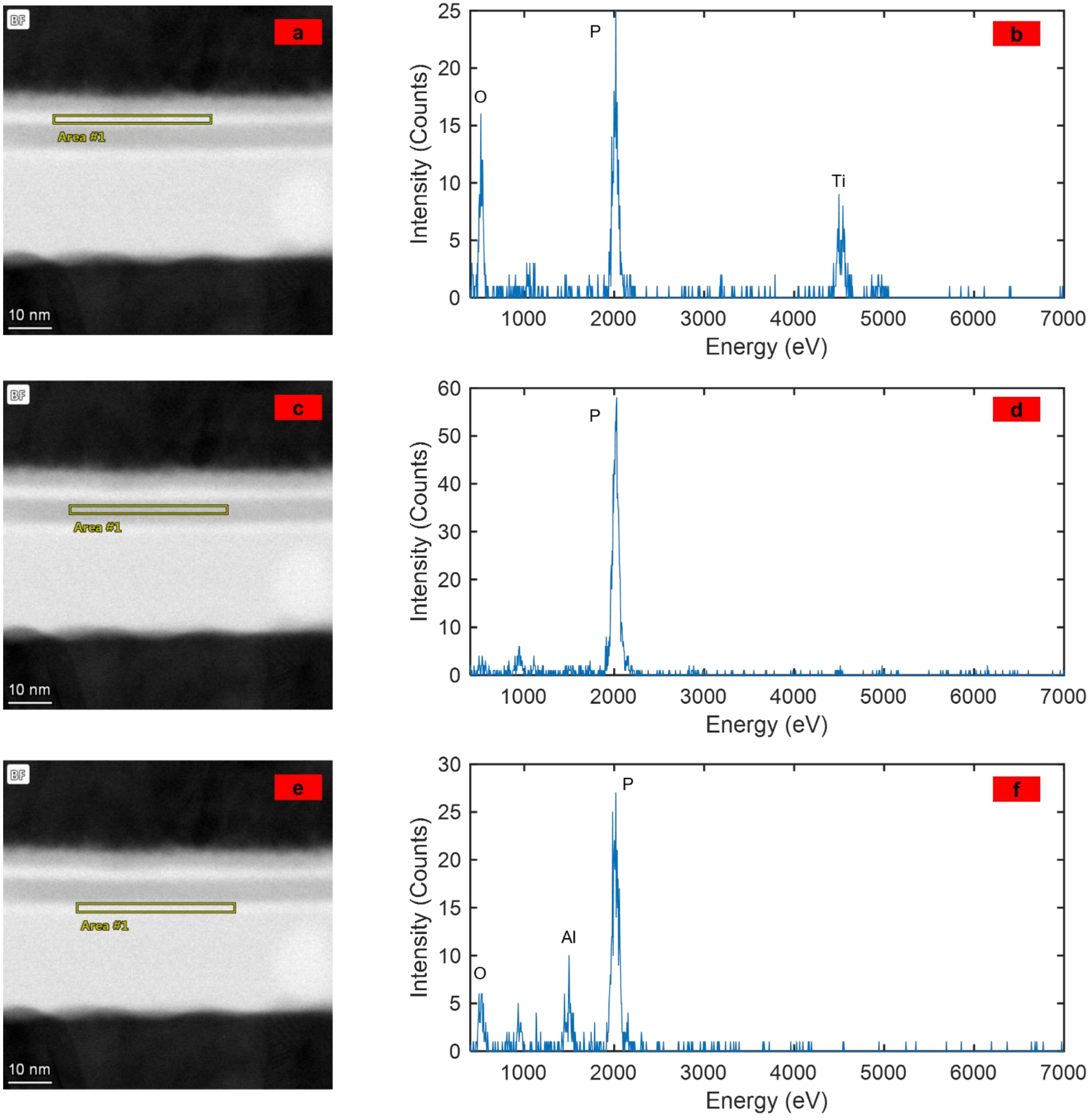

Fig. S4 EDS mapping spectrum

## Supplementary Note 5 | BP-FET parameter library and I-V benchmark plots

Table SI Model parameters library for simulations of the BP-FETs

| | $Y\mu_n$ ($cm^2V^{-1}s^{-1}$) | $YN_{it,e}$ ($cm^{-2}$) | $Y\mu_p$ ($cm^2V^{-1}s^{-1}$) | $YN_{it,h}$ ($cm^{-2}$) | $Y\Phi_t$ (eV) | $Y\varphi_a$ (eV) | $Y\Phi_a$ (eV) | $\sigma_s$ (V) |
|---|---|---|---|---|---|---|---|---|
| Dev#1-2 | | | | | | | | |
| 1 | 6.00E+00 | 2.09E+13 | 1.34E+02 | 5.81E+12 | 3.62E-01 | 1.01E+00 | 5.68E-01 | 7.40E-01 |
| 2 | 2.93E+00 | 7.72E+12 | 4.00E+02 | 9.89E+12 | 5.27E-01 | 2.09E+00 | 7.20E-01 | * |
| 3 | * | * | * | * | 3.85E-01 | 7.00E-01 | 5.66E-01 | * |
| Dev#2-3 | | | | | | | | |
| 1 | 6.01E+00 | 1.94E+13 | 8.77E+01 | 5.81E+12 | 4.67E-01 | 1.32E+00 | 6.14E-01 | 4.31E-01 |
| 2 | 1.77E+00 | 6.15E+12 | 4.01E+02 | 8.02E+12 | 5.46E-01 | 5.30E+00 | 1.30E+00 | * |
| 3 | * | * | * | * | 2.48E-01 | 1.56E+00 | 8.37E-01 | * |
| Dev#3-4 | | | | | | | | |
| 1 | 6.00E+00 | 2.02E+13 | 1.27E+02 | 5.71E+12 | 2.70E-01 | -2.03E-01 | 5.31E-01 | 6.12E-01 |
| 2 | 2.84E+00 | 7.48E+12 | 4.00E+02 | 9.93E+12 | 5.09E-01 | 1.58E+00 | 5.38E-01 | * |
| 3 | * | * | * | * | 3.61E-01 | 9.85E-01 | 4.35E-01 | * |
| Dev#4-5 | | | | | | | | |
| 1 | 6.01E+00 | 1.84E+13 | 7.63E+01 | 4.92E+12 | 2.45E-01 | 6.26E-01 | 1.08E+00 | 1.26E+00 |
| 2 | 1.49E+00 | 6.22E+12 | 4.00E+02 | 7.19E+12 | 4.74E-01 | 5.55E+00 | 2.17E+00 | * |
| 3 | * | * | * | * | 2.42E-01 | -2.35E-01 | 9.61E-01 | * |
| Dev#5-6 | | | | | | | | |
| 1 | 9.86E+00 | 1.64E+13 | 1.13E+02 | 5.32E+12 | 2.64E-01 | 4.98E-01 | 5.50E-01 | 3.07E-01 |
| 2 | 3.66E+00 | 9.12E+12 | 5.81E+02 | 8.02E+12 | 3.88E-01 | -1.00E+00 | 6.17E+00 | * |
| 3 | * | * | * | * | 3.59E-01 | -9.98E-01 | 3.09E+00 | * |
| Dev#1-3 | | | | | | | | |
| 1 | 1.00E+01 | 2.06E+13 | 1.25E+02 | 5.81E+12 | 4.64E-01 | 9.93E-01 | 5.89E-01 | 4.03E-01 |
| 2 | 3.32E+00 | 4.53E+12 | 4.67E+02 | 9.95E+12 | 5.72E-01 | 3.94E+00 | 1.04E+00 | * |
| 3 | * | * | * | * | 2.34E-01 | 1.21E+00 | 5.73E-01 | * |
| Dev#2-4 | | | | | | | | |
| 1 | 9.72E+00 | 2.08E+13 | 1.43E+02 | 5.37E+12 | 3.35E-01 | 7.16E-01 | 4.94E-01 | 6.56E-01 |
| 2 | 3.29E+00 | 6.86E+12 | 4.33E+02 | 7.07E+12 | 4.73E-01 | 2.46E+00 | 1.25E+00 | * |
| 3 | * | * | * | * | 2.92E-01 | -6.78E-02 | 8.00E-01 | * |
| Dev#3-5 | | | | | | | | |
| 1 | 6.00E+00 | 1.98E+13 | 1.39E+02 | 5.79E+12 | 3.25E-01 | 5.42E-01 | 5.08E-01 | 3.05E-01 |
| 2 | 2.80E+00 | 6.23E+12 | 4.73E+02 | 9.02E+12 | 4.79E-01 | 1.63E+00 | 7.70E-01 | * |
| 3 | * | * | * | * | 2.67E-01 | 5.55E-01 | 2.74E-01 | * |
| Dev#4-6 | | | | | | | | |
| 1 | 6.85E+00 | 2.01E+13 | 1.16E+02 | 5.51E+12 | 3.39E-01 | 7.44E-01 | 4.91E-01 | 1.76E-01 |
| 2 | 3.22E+00 | 7.19E+12 | 6.00E+02 | 8.90E+12 | 4.49E-01 | 2.89E+00 | 1.92E+00 | * |
| 3 | * | * | * | * | 3.54E-01 | -4.00E-01 | 6.47E-01 | * |
| Dev#1-4 | | | | | | | | |
| 1 | 9.09E+00 | 1.87E+13 | 1.51E+02 | 5.81E+12 | 4.62E-01 | 8.89E-01 | 4.69E-01 | 1.35E+00 |
| 2 | 5.60E+00 | 5.50E+12 | 4.39E+02 | 7.11E+12 | 5.80E-01 | 3.88E+00 | 9.83E-01 | * |
| 3 | * | * | * | * | 1.89E-01 | -2.49E-01 | 8.32E-01 | * |
| Dev#2-5 | | | | | | | | |
| 1 | 9.53E+00 | 2.08E+13 | 1.30E+02 | 3.97E+12 | 2.32E-01 | 7.51E-01 | 4.82E-01 | 4.26E-01 |
| 2 | 3.76E+00 | 7.55E+12 | 6.00E+02 | 8.78E+12 | 4.38E-01 | 1.48E+00 | 1.99E+00 | * |
| 3 | * | * | * | * | 3.82E-01 | -1.84E-01 | 5.66E-01 | * |
| Dev#3-6 | | | | | | | | |
| 1 | 6.01E+00 | 1.81E+13 | 1.58E+02 | 5.74E+12 | 3.68E-01 | 6.81E-01 | 4.41E-01 | 1.37E-01 |
| 2 | 4.30E+00 | 6.46E+12 | 6.00E+02 | 8.84E+12 | 4.66E-01 | 2.21E+00 | 1.16E+00 | * |
| 3 | * | * | * | * | 2.94E-01 | 4.49E-01 | 2.56E-01 | * |
| Dev#1-5 | | | | | | | | |
| 1 | 9.98E+00 | 1.76E+13 | 1.37E+02 | 3.50E+12 | 2.47E-01 | 8.36E-01 | 4.77E-01 | 3.43E-01 |
| 2 | 5.25E+00 | 5.03E+12 | 6.00E+02 | 7.89E+12 | 4.56E-01 | 2.92E+00 | 1.34E+00 | * |
| 3 | * | * | * | * | 1.94E-01 | -4.18E-01 | 4.51E-01 | * |
| Dev#2-6 | | | | | | | | |
| 1 | 1.00E+01 | 2.07E+13 | 1.67E+02 | 4.52E+12 | 2.94E-01 | 7.38E-01 | 4.03E-01 | 2.27E-01 |
| 2 | 5.57E+00 | 8.27E+12 | 6.00E+02 | 9.13E+12 | 4.46E-01 | 1.30E+00 | 1.86E+00 | * |
| 3 | * | * | * | * | 4.34E-01 | -2.47E-01 | 6.09E-01 | * |
| Dev#1-6 | | | | | | | | |
| 1 | 6.00E+00 | 1.92E+13 | 1.50E+02 | 3.54E+12 | 3.05E-01 | 7.37E-01 | 3.83E-01 | 2.23E-01 |
| 2 | 5.60E+00 | 6.09E+12 | 5.99E+02 | 8.67E+12 | 4.60E-01 | 2.21E+00 | 1.28E+00 | * |
| 3 | * | * | * | * | 3.71E-01 | -3.43E-02 | 3.22E-01 | * |

Simulated output/transfer curves in linear/logarithm scale benchmarked with experimental data are shown below

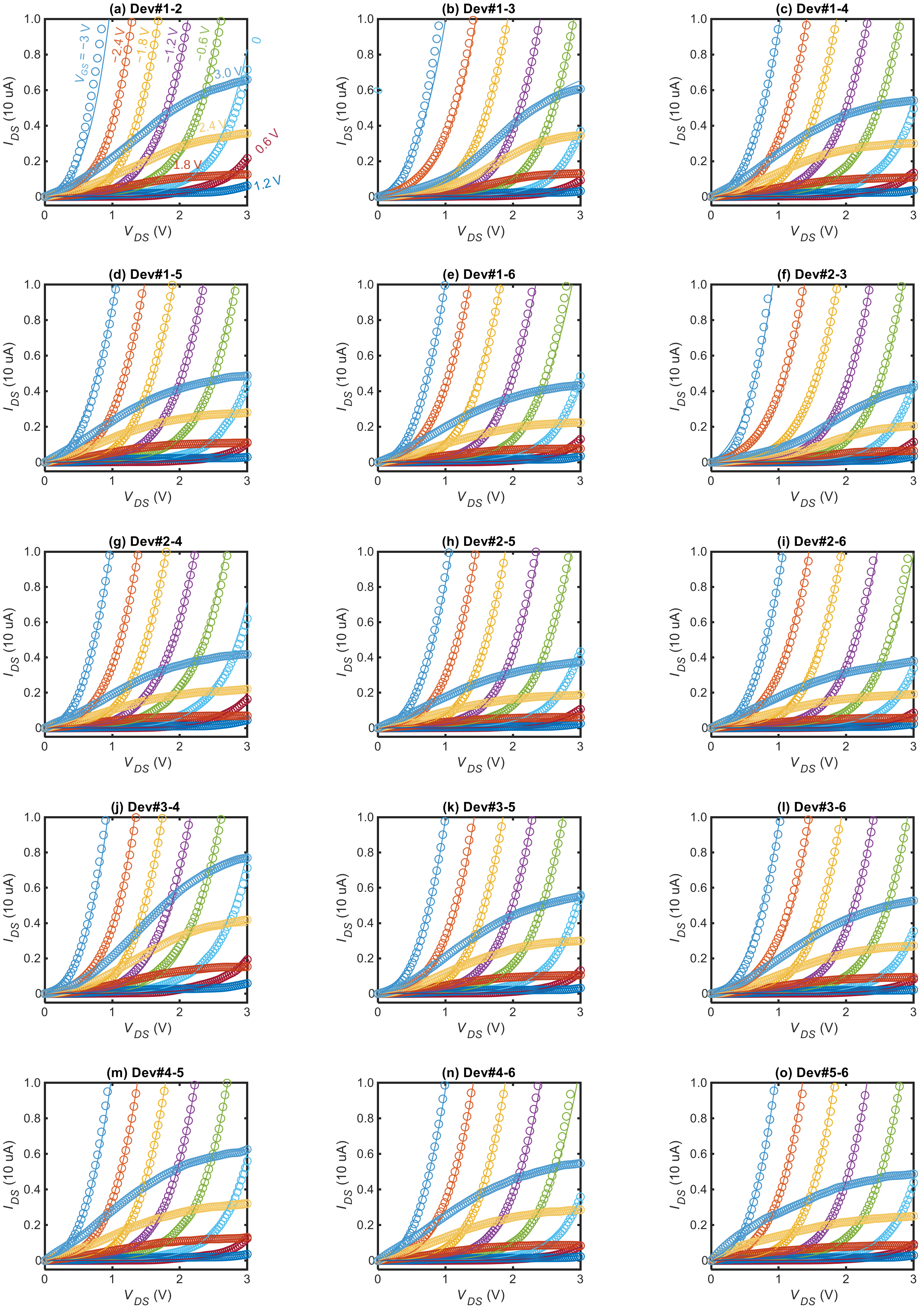

Fig. S5 Simulated output curves (lines) benchmarked with experimental data (circles) in linear scale.

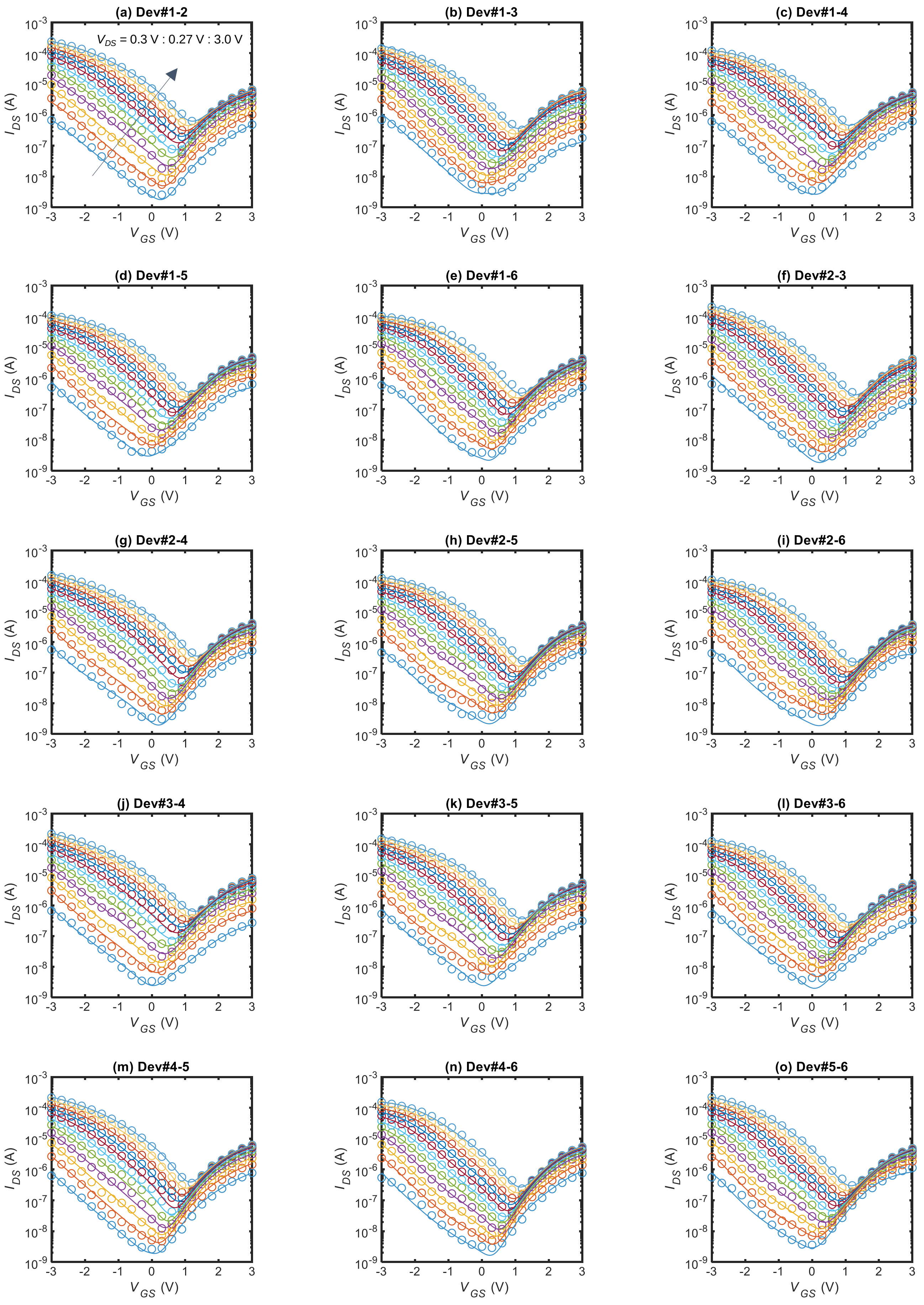

Fig. S6 Simulated transfer curves (lines) benchmarked with experimental data (circles) in logarithm scale.

# Supplementary Note 6 | BP-based ATIQ circuits: Simulation Parameters

Table SII Model parameters library for simulations of the BP-FETs

| | $Y\mu_n$ ($cm^2V^{-1}s^{-1}$) | $YN_{trp,e}$ ($cm^{-2}$) | $Y\mu_p$ ($cm^2V^{-1}s^{-1}$) | $YN_{trp,h}$ ($cm^{-2}$) | $Y\Phi_t$ (eV) | $Y\varphi_a$ (eV) | $Y\Phi_a$ (eV) | $\sigma_s$ (V) |
|---|---|---|---|---|---|---|---|---|
| Dev#3-1 | | | | | | | | |
| 1 | 5.78E+01 | 2.54E+12 | 2.97E+01 | 5.58E+12 | 4.53E-01 | 1.68E-01 | 4.13E-01 | 6.69E-01 |
| 2 | 3.31E+01 | 7.28E+12 | 5.70E+01 | 3.83E+12 | 3.59E-01 | -1.91E+00 | 1.09E-02 | * |
| 3 | * | * | * | * | 4.77E-01 | * | 5.51E+00 | * |
| Dev#3-2 | | | | | | | | |
| 1 | 8.20E+01 | 2.35E+12 | 5.90E+01 | 7.84E+12 | 4.37E-01 | -5.52E+00 | 1.46E+01 | 5.37E-01 |
| 2 | 4.91E+01 | 8.65E+12 | 1.85E-01 | 1.73E+12 | 3.43E-01 | 4.38E+00 | 6.53E-01 | * |
| 3 | * | * | * | * | 4.48E-01 | * | 6.41E+00 | * |
| Dev#3-4 | | | | | | | | |
| 1 | 6.49E+01 | 5.27E+12 | 4.90E+01 | 1.07E+13 | 6.03E-01 | 1.00E-01 | 3.39E-01 | 4.57E-01 |
| 2 | 3.35E+01 | 8.63E+12 | 1.50E+00 | 2.48E+11 | 3.79E-01 | -1.57E+00 | 2.12E-03 | * |
| 3 | * | * | * | * | 4.68E-01 | 8.74E-01 | 6.55E+00 | * |
| Dev#5-3 | | | | | | | | |
| 1 | 2.85E+02 | 2.96E+12 | 1.51E+02 | 1.00E+13 | 4.47E-01 | -2.84E+00 | 2.11E+00 | 9.91E-01 |
| 2 | 2.97E+02 | 9.06E+12 | 4.26E+00 | 2.35E+12 | 3.00E-01 | 3.03E+00 | 1.58E+00 | * |
| 3 | * | * | * | * | 2.54E-01 | 5.42E+00 | * | * |
| Dev#6-5 | | | | | | | | |
| 1 | 2.97E+02 | 1.19E+13 | 1.17E+02 | 1.56E+13 | 1.04E+00 | 1.55E+00 | 5.14E-01 | 7.95E-01 |
| 2 | 4.90E+01 | 9.25E+12 | 6.00E+02 | 2.78E+12 | 4.89E-01 | 1.87E+00 | 2.36E+00 | * |
| 3 | * | * | * | * | 3.32E-01 | -1.67E-01 | * | * |

## Supplementary Note 7 | $MoS_2$ FET: Simulation Parameters

Table SIII Model parameters library for simulations of the $MoS_2$-FETs

| T0 | $Y\mu_n$ (cm$^2$V$^{-1}$s$^{-1}$) | $YN_{trp,e}$ (cm$^{-2}$) | $Y\mu_p$ (cm$^2$V$^{-1}$s$^{-1}$) | $YN_{trp,h}$ (cm$^{-2}$) | $Y\Phi_t$ (eV) | $Y\varphi_a$ (eV) | $Y\Phi_a$ (eV) | $\sigma_s$ (V) |
|---|---|---|---|---|---|---|---|---|
| 1 | 1.86E+00 | 4.04E+12 | * | 3.48E+12 | 2.64E-01 | -3.93E-01 | 2.54E-01 | 1.00E+00 |
| 2 | 1.86E+00 | 2.97E+12 | * | 9.19E+12 | 4.99E-01 | 1.93E-01 | 6.10E-01 | * |
| 3 | * | * | * | * | 2.64E-01 | 1.40E-01 | 5.73E-01 | * |